\documentclass[twocolumn]{aastex631}

\usepackage{amsmath}
\usepackage{float}
\usepackage{xcolor}

\begin{document}

\title{Chromaticity-Optimized Antenna Design and Bayesian Foreground Validation for the CANTAR Global 21 cm Experiment}

\author{Michelle Mora}
\affiliation{FACom, Instituto de Física - FCEN, Universidad de Antioquia \\
Medellín, Colombia}

\author{Germán Chaparro}
\affiliation{FACom, Instituto de Física - FCEN, Universidad de Antioquia \\
Medellín, Colombia}
\correspondingauthor{Germán Chaparro}
\email{german.chaparro@udea.edu.co}

\author{Juan D. Guerrero}
\affiliation{FACom, Instituto de Física - FCEN, Universidad de Antioquia \\
Medellín, Colombia}

\author{Catalina Alzate}
\affiliation{FACom, Instituto de Física - FCEN, Universidad de Antioquia \\
Medellín, Colombia}

\author{Juan P. Urrego}
\affiliation{FACom, Instituto de Física - FCEN, Universidad de Antioquia \\
Medellín, Colombia}

\author{Jimena Giraldo}
\affiliation{FACom, Instituto de Física - FCEN, Universidad de Antioquia \\
Medellín, Colombia}

\author{Pablo Cuartas-Restrepo}
\affiliation{FACom, Instituto de Física - FCEN, Universidad de Antioquia \\
Medellín, Colombia}

\author{Julian Rodriguez-Ferreira}
\affiliation{Grupo de investigación en control, electrónica, modelado y simulación CEMOS. Escuela de ingenierías eléctrica, electrónica y telecomunicaciones. Universidad Industrial de Santander \\
Bucaramanga, Colombia}

\author{Oscar Restrepo}
\affiliation{Grupo SiAMo, Vicerrectoría de Investigación, Universidad ECCI \\
Bogotá, Colombia}



\begin{abstract}

Detecting the global 21\,cm signal from the epoch of reionization remains a major observational challenge due to bright foregrounds and instrumental systematics. As part of the Colombian Antarctic Telescopes for 21\,cm Absorption during Reionization (CANTAR) initiative, we present a simulation and analysis framework to evaluate antenna chromaticity, optimize instrument design, and assess site suitability for global 21\,cm experiments. Using frequency-dependent beam models and Haslam-based sky maps, we compute dynamic spectra for the EDGES blade dipole and a set of dipole and novel monopole antennas optimized via particle swarm optimization. The optimized designs exhibit improved spectral smoothness compared to EDGES, particularly in the 70-120\,MHz range. We also evaluate latitude-dependent sky brightness and identify mid-latitude sites ($-40^\circ$ to $+5^\circ$) as optimal for foreground suppression.  {We apply Bayesian inference together with posterior predictive model validation to the publicly released EDGES data, assessing statistical consistency rather than hypothesis testing or model comparison. We find that physically motivated foreground and ionospheric models are statistically consistent with the data only when a 21 cm absorption feature is excluded. } From the validated posterior, we generate a statistically validated ensemble of foreground corrections for use in beam–sky simulations. These results support a two-phase strategy for CANTAR: Antarctic deployments for calibration and testing, and future science operations at mid-latitude sites. Our framework provides a validated path toward robust foreground modeling, antenna design, and systematics control for global 21\,cm signal detection.

\end{abstract}



\section{Introduction} \label{sec:intro}

The redshifted 21 cm hyperfine transition of neutral hydrogen is a powerful probe of the early Universe, enabling direct constraints on the thermal and ionization history of the intergalactic medium (IGM) during cosmic dawn and the Epoch of Reionization \citep{furlanetto2006cosmology,Pritchard, liu2020data}. Among the most promising observables is the global 21 cm signal: the sky-averaged brightness temperature as a function of frequency, which encodes information about the formation of the first stars and the timing of heating and reionization.

Global 21 cm experiments must contend with foregrounds that are $10^4$ to $10^5$ times brighter than the expected cosmological signal, primarily synchrotron radiation from the Milky Way \citep{Rogers_2008,mozdzen}. In addition, instrumental systematics—especially beam chromaticity, impedance mismatch, and ionospheric effects—introduce spectral features that can mimic or obscure a real signal \citep{veda,bernardi,hibbard2020modeling}. Proposed projects such as DAPPER have gone as far as to propose the installation of an instrument in the far side lunar environment to overcome both RFI and the ionospheric foreground issues \citep{burns2020}.

The first reported detection of a global 21\,cm absorption feature was made by the Experiment to Detect the Global Epoch of Reionization Signature (EDGES) team \citep{bowman}. The signal, centered near 78\,MHz, exhibited an amplitude more than twice as deep as predicted by standard cosmological models. Such a strong absorption would require the primordial gas to be significantly colder than expected, potentially due to exotic physics such as baryon–dark matter interactions \citep[]{cita4} or {or an enhanced radio background exceeding the CMB at the relevant redshifts}. In particular, explaining the EDGES result in this framework would require dark matter particles with masses several times smaller than typically assumed. This tension has led to questions about the robustness of the EDGES measurement, including the possibility of unmodeled instrumental systematics such as antenna chromaticity \citep[]{veda}, as well as the lack of a rigorous statistical validation of the proposed foreground and signal model. 

{Several Bayesian reanalyses have raised concerns regarding instrumental systematics and foreground model degeneracy \citep{sims,hills2018}, while others focus on systematics mitigation for future experiments \citep{murray,kirkham}.} Moreover, recent SARAS 3 \citep{saras} results have {found no evidence for an absorption feature of similar depth to that reported by EDGES, despite claiming comparable instrumental sensitivity.} Other efforts have applied different techniques to confront foreground contamination challenges \citep{scihi,amiri2022,liu2022, mondal2023, dasgupta2023}: the PRIZM \citep{prizm} and MIST \citep{mist,monsalve} instruments seek to improve sensitivity through optimized site selection and system design.

Recent efforts have focused on mitigating beam chromaticity both through hardware optimization and joint calibration methods \citep{hibbard2020modeling,anstey2021general,mahesh2021validation,sun2024}
. The REACH experiment, for example, uses a Bayesian framework that simultaneously fits the 21 cm signal, foregrounds, and chromatic beam response \citep{reach,reach2}. Computer-assisted antenna design optimization has only recently been applied to the problem of chromaticity mitigation \citep{restrepo}.

In this work, we present an integrated simulation and analysis framework to support the design and interpretation of single-element global 21 cm experiments. This work was developed within the context of the Colombian Antarctic Telescopes for 21 cm Absorption during Reionization (CANTAR) initiative \citep{cantar}, which aims to develop and validate a full pipeline for global 21 cm signal detection through optimized instrumentation, statistical modeling, and site characterization. We simulate beam-weighted sky spectra for several antenna configurations, including the EDGES blade dipole, two blade dipoles optimized via particle swarm optimization (PSO) originally developed in \citet{restrepo}, and three novel monopole designs intended for site testing and foreground model validation. These simulations use frequency-dependent antenna beam models and sky brightness distributions extrapolated from the 408 MHz Haslam survey.

We also assess the dependence of integrated sky spectra on observer latitude, identifying optimal geographic locations that minimize Galactic foreground contamination. Finally, we apply a Bayesian inference framework to the publicly released EDGES data to evaluate the robustness of the reported absorption feature, {emphasizing on posterior predictive model validation, in the sense of assessing whether a given generative model is statistically consistent with the observed data structure and variance, rather than whether it is merely preferred relative to alternative models.} Specifically, we test the foreground and ionospheric model proposed by \cite{hills2018} and show that it can reproduce the observed spectrum only when significantly larger observational uncertainties are assumed. The best-fitting models do not require a 21 cm absorption signal, and instead reflect a relaxation of the assumed error model to account for unexplained residual structure. These results cast doubt on the statistical interpretation of the EDGES detection under standard assumptions. We incorporate the resulting posterior distributions into our beam–sky simulations to generate a statistically validated ensemble of radio foregrounds for ongoing CANTAR design and site studies.

This paper is structured as follows. In Section 2, we describe our simulation pipeline, antenna modeling, and statistical methodology. Section 3 presents results on beam chromaticity, Bayesian model validation, and observational site optimization. We conclude in Section 4 with conclusions and a discussion of implications for future global 21 cm detection efforts.

\begin{figure*}[t]
    \centering
    \gridline{
        \fig{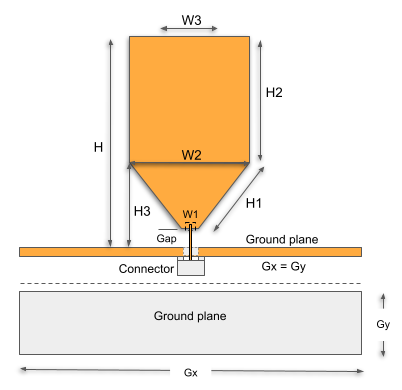}{0.32\textwidth}{(a)}
        \fig{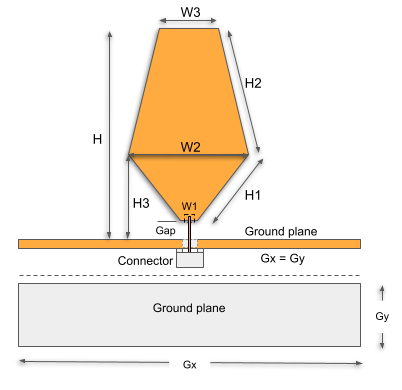}{0.32\textwidth}{(b)}
        \fig{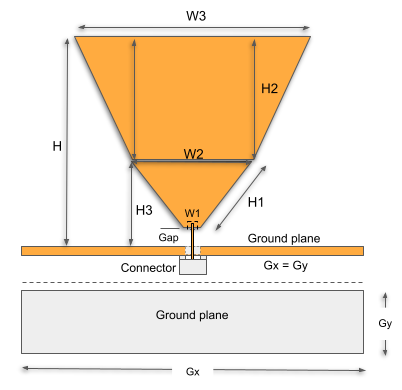}{0.32\textwidth}{(c)}
    }
    \caption{
        Geometry of the three PSO-optimized monopole blade antennas proposed for site testing, each mounted over a metallic ground plane. All designs share the same values of $W_2$, $H_2$, and $H_3$, but differ in the width of the upper blade section $W_3$, which controls impedance bandwidth.
        \textbf{(a)} Monopole-1: baseline configuration with $W_2 = W_3$.
        \textbf{(b)} Monopole-2: wingless variant with $W_3 < W_2$, optimized for narrower high-frequency response.
        \textbf{(c)} Monopole-3: winged variant with $W_3 > W_2$, designed to enhance low-frequency coupling.
        These designs were selected to explore how blade asymmetry impacts chromaticity and suitability for foreground modeling in site tests.
    }
    \label{Fig:geometry_monopole_antennas}
\end{figure*}

\section{Methods} \label{sec:Met}

The simulations and analysis presented in this work were developed as part of the CANTAR (Colombian Antarctic Telescopes for 21 cm Absorption during Reionization) collaboration \citep{cantar}. Our goal is to evaluate chromatic antenna response, simulate beam-weighted sky spectra, and statistically validate foreground corrections relevant to global 21 cm signal detection. All code and data products are publicly available at \url{https://doi.org/10.5281/zenodo.18136149} \citep{zenodo}.

\subsection{Physical Background and Signal Model}
\subsubsection{21 cm Cosmological Line Profile}

The 21\,cm line arises from the hyperfine transition between the singlet and triplet states of neutral hydrogen’s ground state, corresponding to a rest-frame wavelength of 21.11\,cm (or 1420.41\,MHz). This splitting is caused by the interaction between the magnetic moments of the electron and the proton. The transition occurs when the spins of the electron and proton flip from parallel to antiparallel, releasing a photon. The associated energy difference corresponds to a temperature known as the spin temperature, defined as $T_s = 0.0068\,\mathrm{K}$ \citep[]{cita1}.

The observed brightness temperature of the 21\,cm line along a given line of sight depends on the contrast between the spin temperature $T_s$ and the cosmic microwave background (CMB) temperature $T_{\mathrm{cmb}}$, and is modulated by the optical depth $\tau$ of the neutral hydrogen. This can be expressed as:

\begin{equation}
    T_{21} = \frac{T_s - T_{\mathrm{cmb}}}{1 + z} \left(1 - e^{-\tau}\right),
    \label{eq:temp_21}
\end{equation}

where $z$ is the redshift of the emitting region. Around $z \sim 17$, standard cosmological models predict an absorption feature with a depth of approximately $0.2\,\mathrm{K}$ \citep[]{cita6}.

To detect such a faint signal, one must account for the much brighter radio emission from Galactic and terrestrial sources. The total observed sky temperature, $T_{\mathrm{sky}}$, includes the cosmological signal $T_{21}$ and the foreground temperature $T_F$, such that:
\[
T_{\mathrm{sky}}(\nu) = T_{21}(\nu) + T_F(\nu).
\]
\begin{figure*}
    \centering
    \includegraphics[width=0.8\textwidth]{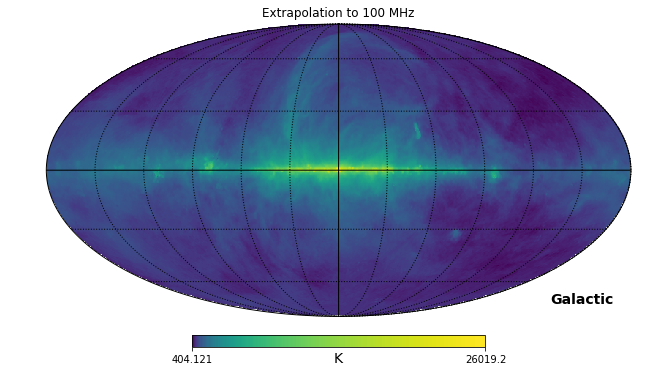}
    \caption{\citet{HaslamMap} 408\,MHz sky map extrapolated to 100\,MHz assuming a spectral index $\alpha = 2.5$, in Galactic coordinates. The interpolation to each frequency serves as the foreground model in our beam–sky convolution simulations.
}
    \label{fig:Haslam 100}
 \end{figure*}

The EDGES collaboration modeled the absorption profile using an empirical, flattened Gaussian function \citep{bowman}. Although not physically motivated, this form yields a convenient phenomenological fit to the data. The model is defined as,

\begin{equation}
    T_{21}(\nu) = -A \left( \frac{1 - e^{-\tau e^B}}{1 - e^{-\tau}} \right),
    \label{eq:T_21}
\end{equation}

where

\begin{equation}
    B = \frac{4 (\nu - \nu_0)^2}{w^2} \log \left[-\frac{1}{\tau} \log \left( \frac{1 + e^{-\tau}}{2} \right) \right],
    \label{eq:B}
\end{equation}

and $A$ is the absorption amplitude, $\nu_0$ is the central frequency, $w$ is the full width at half maximum (FWHM), and $\tau$ is the flattening factor. As noted by \citet{bowman}, this model does not describe the physical origin of the absorption but serves as a flexible parametric form for fitting broad features in the data.

\subsubsection{Sky foreground simulation}
\label{fore}
Our evaluation of antenna performance is based on simulating beam-weighted observations of a synthetic radio sky. As a baseline sky model for $T_F$, we used the all-sky 408 MHz map from \citet{HaslamMap}, which is dominated by Galactic synchrotron emission. To extend this map across our frequency range of interest, we applied a power-law spectral extrapolation consistent with synchrotron emission, yielding a frequency-dependent model of the diffuse radio foreground.
\begin{equation}\label{eq:Textrapolation}
T_\text{sky}=T_0 \Big(\frac{\nu}{\nu_c}\Big)^{-\alpha} \ .
\end{equation}

Here $\nu$ is the frequency, $T_0$ is the sky temperature at $\nu_c = 408$ MHz from the \citet{HaslamMap} map, and we took $\alpha = 2.5$ as a naive, initial power-law spectral index \citep{Rogers_2008} that roughly corresponds to the background synchrotron Galactic emission. We assess the role of variations of this index in Sect.~\ref{stoch}.

To map the sky models to the beam simulation data, we created a HEALPix \footnote{http://healpix.sourceforge.net} array based on Galactic coordinates, using the Python library HEALpy \citep{Healpy1, Healpy2}. This allowed us to make a one-to-one correspondence from the antenna gain array to the Healpy pixels in the \citet{HaslamMap} map, thus optimizing the computation time. We then extrapolated to our working frequencies using equation~(\ref{eq:Textrapolation}). Fig.~\ref{fig:Haslam 100} shows the sky map in Galactic coordinates extrapolated to a frequency of 100 MHz. 

The foreground component $T_F$ should include contributions from synchrotron emission in the Galaxy and thermal emission/absorption from the ionosphere. These must be modeled and subtracted to isolate the cosmological signal. A functional form for $T_F(\nu)$ is given by \citet{hills2018} and is discussed in detail in Section~\ref{fore}.

\subsection{Antenna design and beam simulations}

The optimized antennas modelled here are based on a dipole antenna design made of crossed dipoles connected to a ground plane similar to the EDGES antenna design, although that antenna had a custom, manual optimization design process. In contrast, \citet{restrepo} developed a Particle Swarm Optimization (PSO) scheme for improving antenna chromaticity in the frequency range of interest (50 to 150 MHz). First, we consider two PSO-aided dipole antenna designs proposed for 21 cm cosmology in \citet{restrepo}: a rectangular blade design and a "bowtie" design, which for convenience will hereafter be referred to as opt-blade and opt-bowtie.

\begin{figure}[t]
            \centering
            \includegraphics[scale=0.3]{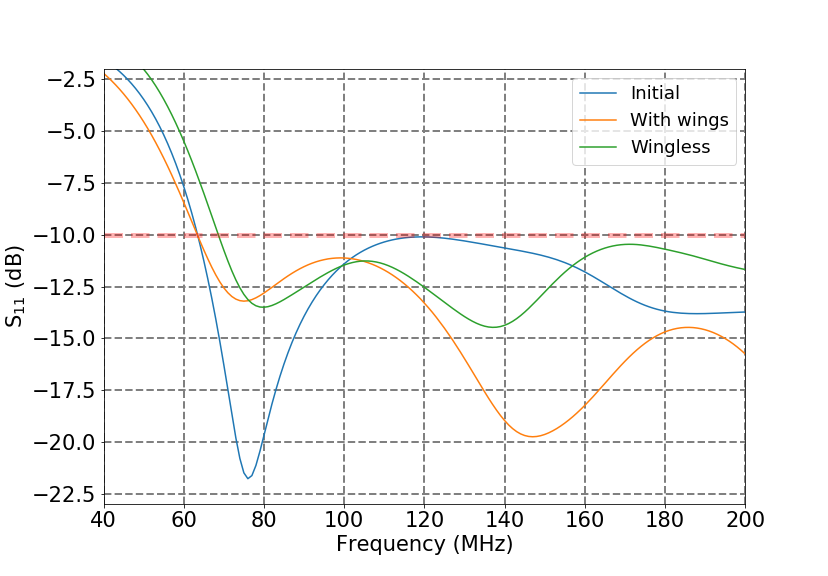}
            \caption{$S_{11}$ parameter for the three PSO-optimized monopole antenna models. The frequency range of interest for global 21\,cm observations is indicated, and all designs achieve reflection coefficients below $-10$\,dB across the target band.}
            \label{fig:Optimized_ All_S11}  
    \end{figure}

Additionally, we analyze three novel, low-cost monopole blade antenna designs intended for site testing in both the low- and high-frequency bands relevant to global 21\,cm cosmology. These designs, shown in Fig.~\ref{Fig:geometry_monopole_antennas}, were optimized using particle swarm optimization (PSO) following the approach of \citet{restrepo}, with the objective of maximizing impedance bandwidth and minimizing the reflection coefficient ($S_{11}$) below $-10$\,dB around the desired center frequency. The resulting $S_{11}$ spectra are shown in Fig.~\ref{fig:Optimized_ All_S11}.

The initial geometry is defined by fixed values of $W_2 = 800$\,mm, $H_2 = 1037$\,mm, and $H_3 = 209$\,mm. We then optimize the parameter $W_3$ to further improve the bandwidth while holding the other dimensions constant. The overall monopole height is set to $H = 1250$\,mm, and the rectangular ground plane measures $600 \times 1690$\,mm. In this configuration, $H$ determines the lower limit of the matched frequency band, while $W_1$ and the feed gap define the upper limit.

{Electromagnetic simulations were carried out using a full-wave finite-element solver (HFSS) following the procedure described in \citet{restrepo}. Convergence was verified by refining both the adaptive mesh and the frequency sampling until changes in the simulated beam patterns, return loss, and chromaticity metrics were below the percent level across the band. The simulation domain included a finite metallic ground plane whose dimensions were fixed during the optimization and chosen to be large compared to the antenna size, ensuring that edge effects did not dominate the radiation pattern. All optimized antenna geometries were evaluated using the same ground-plane configuration to enable direct comparison of chromaticity performance. Additional validation of the simulations was provided by measurements of scaled antenna prototypes, which showed good agreement with the simulated S11 and beam patterns across the accessible frequency range.}

These simulations yield a chromatic beam response array $G[\nu, \theta, \phi]$ in alt-azimuth coordinates. These gain patterns are used in the {time-dependent} beam–sky convolution stage of the dynamic spectrum simulations. The gain matrix has the following structure,

\begin{equation}\label{eq:G_Matrix}
    G[\nu,\theta,\phi]=\begin{bmatrix}
    \begin{bmatrix}
    g_{\nu_0}(\theta,\phi)
    \end{bmatrix}\\ 
    \begin{bmatrix}
    g_{\nu_1}(\theta,\phi)
    \end{bmatrix}\\ 
    \vdots \\ 
    \begin{bmatrix}
    g_{\nu_m}(\theta,\phi)
    \end{bmatrix}\\ 
    \end{bmatrix} .
\end{equation}

Here $g_{\nu_n}(\theta,\phi)$ is the beam gain simulation result for the $n$-th frequency in the range, organized in a matrix where the columns represent the azimuth $\phi$ and the rows the altitude $\theta$, both with an angular resolution of 1 degree. The beam gains of the opt-blade and opt-bowtie antennas are reported in \citet{restrepo}.

\begin{figure*}[htbp!]
    \centering
    \gridline{
        \fig{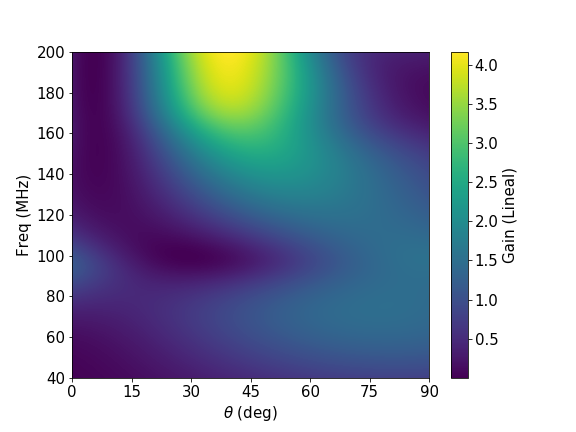}{0.32\textwidth}{(a)}
        \fig{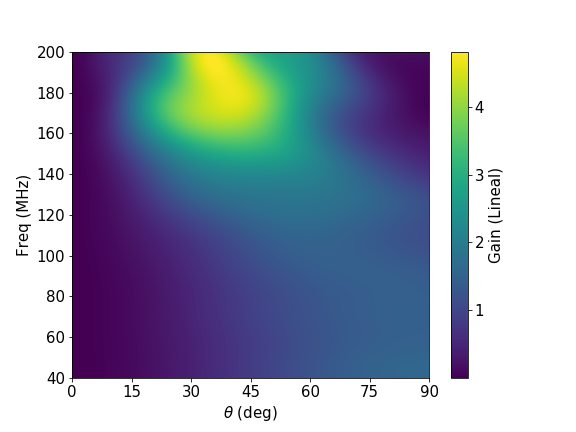}{0.32\textwidth}{(b)}
        \fig{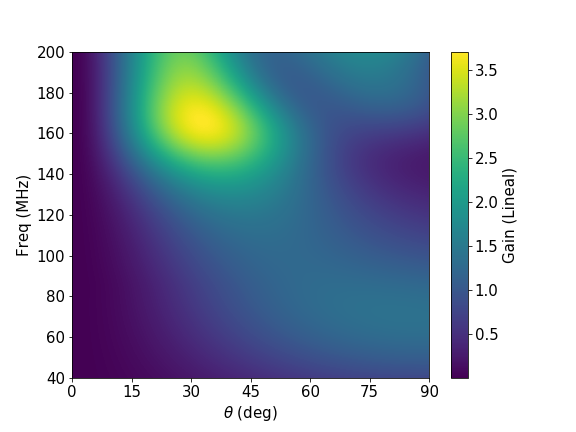}{0.32\textwidth}{(c)}
    }
    \vspace{-0.4cm}
    \gridline{
        \fig{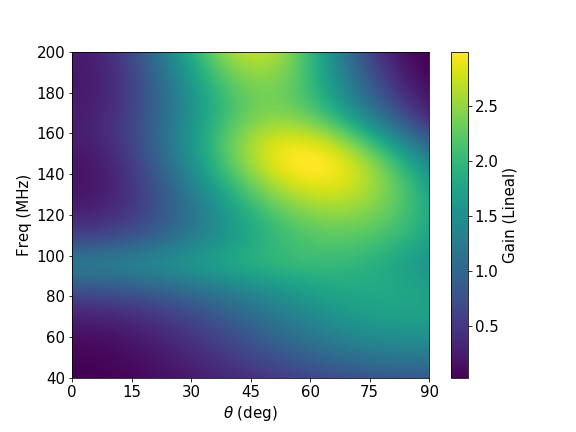}{0.32\textwidth}{(d)}
        \fig{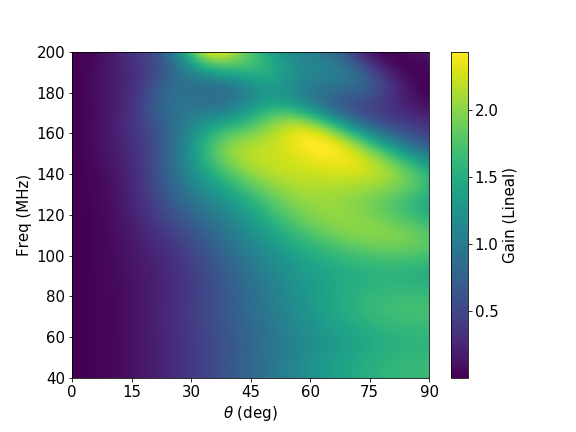}{0.32\textwidth}{(e)}
        \fig{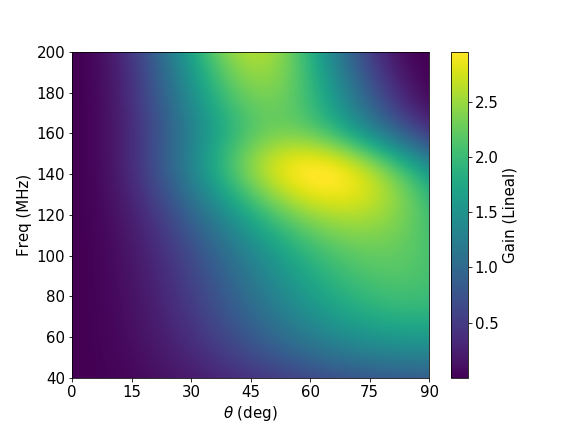}{0.32\textwidth}{(f)}
    }
    \caption{
        ANSYS-HFSS simulated gain patterns for the proposed monopole blade antennas, {with $\theta$ being the altitude, and $\phi$ the azimuth}.
        Top panels: cuts in the $\phi = 90^\circ$ plane.
        Bottom panels: cuts in the $\phi = 0^\circ$ plane.
        These slices illustrate the angular response and beam symmetry across the sky for each design.
    }
    \label{Fig:gain_all_monopole_antennas}
\end{figure*}

\begin{figure*}
    \centering
    \gridline{
        \fig{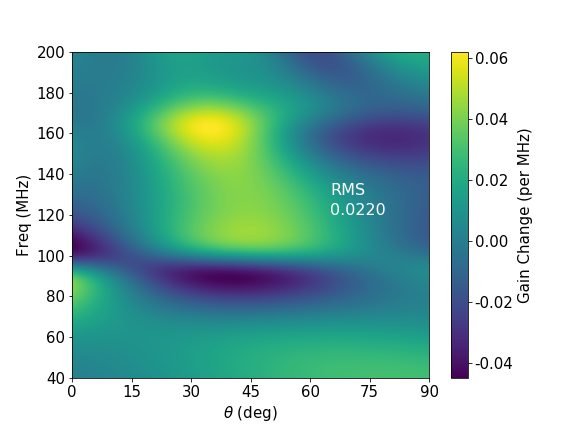}{0.32\textwidth}{(a)}
        \fig{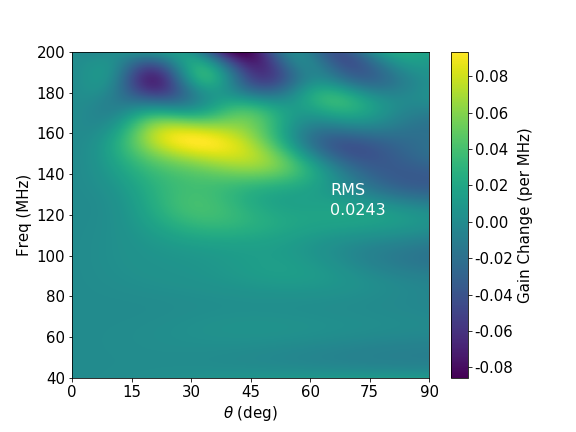}{0.32\textwidth}{(b)}
        \fig{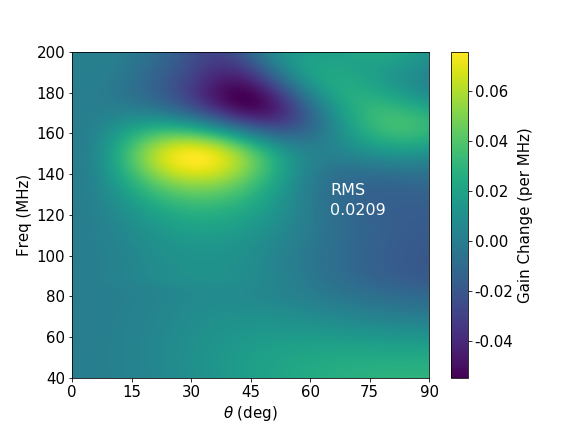}{0.32\textwidth}{(c)}
    }
    \vspace{-0.4cm}
    \gridline{
        \fig{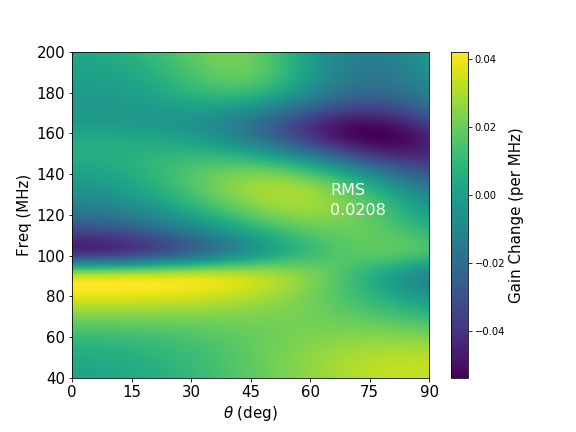}{0.32\textwidth}{(d)}
        \fig{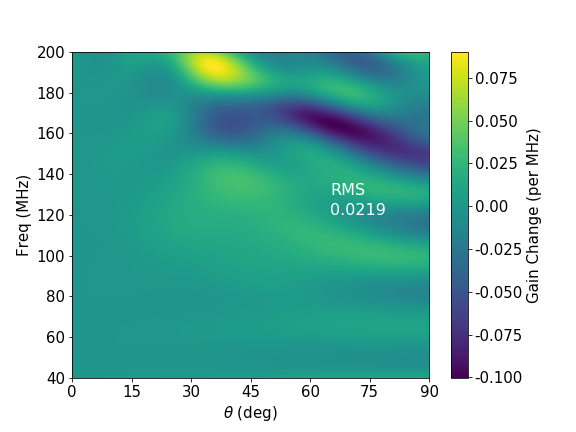}{0.32\textwidth}{(e)}
        \fig{Change_Gain_40_200_MHz_interpole_with_wings_phi_90.png}{0.32\textwidth}{(f)}
    }
    \caption{
        Linear gain gradient for the monopole blade antennas.
        Top panels: $\phi = 90^\circ$ plane.
        Bottom panels: $\phi = 0^\circ$ plane.
        The gradients are a measure of the beam chromaticity across frequency and angle for each design.
    }
    \label{Fig:changegain_all_monopole_antennas}
\end{figure*}

Figure \ref{Fig:gain_all_monopole_antennas} shows cuts of the beam gains of the monopole site testing antennas at $\phi= 0^\circ$ and $\phi= 90^\circ$. Figure \ref{Fig:changegain_all_monopole_antennas} shows the beam gain gradient for each proposed antenna. 

To quantify antenna chromaticity, we compute the root mean square (RMS) of the gain gradient across the frequency band of interest, following standard practice in global 21 cm experiments \citep{mozdzen}. the commonly used benchmark for 21-cm cosmology of $\lesssim0.1$ MHz$^{-1}$, whereas our optimized antennas achieve values $\lesssim0.02$ MHz$^{-1}$. This represents about a factor of five improvement compared to typical performance criteria used in earlier experiments. Explicitly, we define the chromaticity as the RMS of the frequency derivative of antenna gain normalized by the gain itself, and our results demonstrate that these PSO-optimized designs significantly surpass conventional benchmarks.

\subsection{Sky model and beam convolution}

\begin{figure*}[t]
\centering
\includegraphics[width=0.8\textwidth]{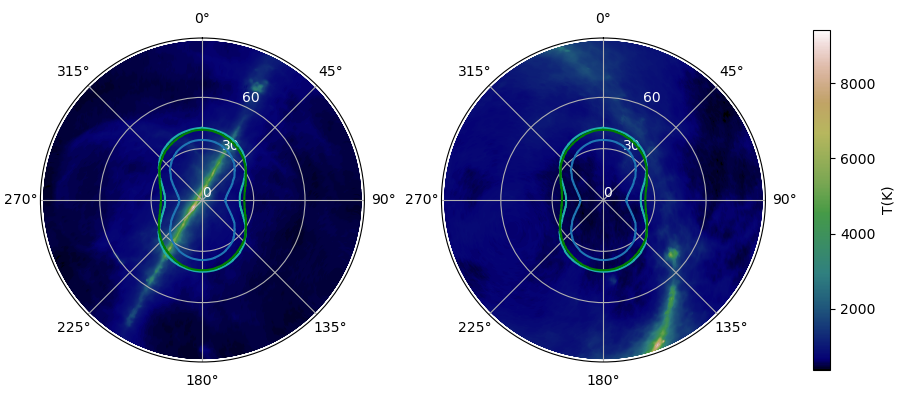}
\caption{Simulated 100MHz sky temperature at the ALMA site ($23.03^\circ$\,S, $67.76^\circ$\,W) on 2021-06-14. Left: 05:00 UTC with Galactic longitude $0^\circ$ at zenith; right: 16:00 UTC with longitude $224^\circ$ at zenith. Colored contours show antenna beams for the opt-blade (sea green), opt-bowtie (green), and EDGES (blue) designs.
}
\label{fig:Sky_beam_gallong} 
\end{figure*}

When creating a model of the sky, it is important to keep in mind that only half of the sky can be observed from any given location at any particular time of day due to the horizon. This means that radio frequencies may show an active sky during certain times when the Galactic centre is above the horizon, while at other times, the sky will be quieter in these frequencies. By understanding this movement and how it affects observations, we can determine how the temperature changes throughout the day. Figure~\ref{fig:Sky_beam_gallong} provides an example of the sky observed at different times from the location of the ALMA (Atacama Millimeter/Sub-mm Array) site. This is reasonably close to a previously proposed observing site for 21 cm cosmology experiments\footnote{MIST Memo 29 \url{http://www.physics.mcgill.ca/mist/memos/MIST_memo_29.pdf}.}.


To evaluate the antenna performance, we obtain dynamic spectra for various times of the day and orientations of the antenna for different pointings in Galactic coordinates \citep{veda}. To make these dynamic spectra, we simulate observations from the ALMA site location ($23.03^{\circ}$ S, $67.76^{\circ}$ W), and from the Murchison Radioastronomy Observatory location ($26.70^{\circ}$ S, $116.67^{\circ}$ E) for the EDGES sky model. Without loss of generality, we choose an arbitrary day (2021-06-14) for the simulation.

We begin by transforming the Healpy pixel Galactic coordinates to horizontal coordinates using the Python package Astropy\footnote{http://www.astropy.org \citep{astropy:2018}}  starting at 00:00 UT. Then, we make a mask removing all pixels below the horizon, representing what the antenna can observe at a given time. We should note here that these simulations do not consider ground reflections because proper ground conductivity can only be achieved by directly measuring it on-site.

Afterwards, for each frequency in the range, we extrapolate the sky temperature from the \citet{HaslamMap} Map using equation~(\ref{eq:Textrapolation}). We then convolve the new sky temperature with the beam pattern of the antenna according to the $G[\nu,\theta,\phi]$ matrix. Thus, we associate the temperature of each pixel with the respective antenna gain of that point in the sky, simulating the brightness temperature of each sky pixel by computing the following integral \citet{Tools},
\begin{equation}\label{eq:Beam Correction}
T_\text{ant}(\nu)=\frac{\int d\Omega \ B(\theta,\phi,\nu)\ T_\text{sky}(\theta,\phi,\nu)}{\int d\Omega \  B(\theta,\phi,\nu)} \ .
\end{equation}

Here $B(\theta,\phi,\nu)$ is the simulated beam pattern of the antenna dependent on frequency $\nu$ and sky pointing $\theta$, $\phi$, and $T_\text{sky}$ is the extrapolated sky temperature from equation \eqref{eq:Textrapolation}. This process yields the instantaneous, beam-corrected simulated sky temperature $T_\text{ant}$ measured by an antenna at 00:00 UTC. {We repeat this calculation for each frequency channel at six-minute intervals throughout the day, thereby constructing the dynamic spectrum for each antenna.}


\subsection{Observation and site latitude strategies}

Accurately identifying optimal observing times is critical for minimizing contamination from bright foreground structures. In particular, the sky is most favorable for global 21 cm observations when the Galactic plane, especially the region between 90$^\circ$ and 270$^\circ$ in Galactic longitude, is below the horizon. Since the antenna beam and pointing configuration determine the relative weight of the zenith region, it is important to track how this region intersects with Galactic emission over time. To assess this, we compute the Galactic coordinates of the zenith throughout a 24-hour sidereal cycle at each observing latitude.

To evaluate the impact of observing latitude on dynamic spectra, we simulated antenna observations at a series of latitudes spaced by 5 degrees, while fixing the longitude at 0 degrees. For each location, we computed the zenith Galactic coordinates over a full sidereal day to determine when the brightest regions of the sky, defined as Galactic longitudes between 90$^\circ$ and 270$^\circ$, were visible. This is motivated by the presence of fewer bright Galactic radio sources in these regions. Specifically, this longitude range avoids the Galactic center and bright emission from prominent Galactic structures such as the Sagittarius and Perseus arms, significantly reducing foreground brightness. 

We then constructed a latitude-dependent time mask that excludes these high-foreground intervals. The final foreground spectra were generated only for unmasked periods, when the zenith was oriented away from the brightest regions of the Galactic plane, as indicated in the dynamic spectral plots by a shaded box. Thus, our masking strategy explicitly targets periods of minimal Galactic contamination, aiming to improve sensitivity to potential cosmological signals.


\subsection{Bayesian framework and statistical validation}

\subsubsection{Foreground correction and ionospheric effects}
\label{stoch}

The foreground interpolation method shown in Sect.\ref{fore} does not consider the effects of spectral index variations at low frequencies due to the onset of Synchrotron Self-Absorption (SSA), nor for radiative transfer effects of the foreground radiation through the ionosphere. To incorporate these effects, we modify the integrated sky-beam spectra after simulation by applying a physically motivated model of the observed sky temperature $T_e$. Following \citet{hills2018}, we treat spectral index curvature as a correction to the input spectrum prior to ionospheric transfer. The resulting expression for the modified foreground brightness temperature scaled to a reference frequency $\nu_\ast$,
{\small
\begin{equation}
T_\mathrm{sky} = b_0 \left( \frac{\nu}{\nu_\ast} \right)^{\alpha'(\nu)} e^{-b_3 \left( \frac{\nu}{\nu_\ast} \right)^{-2}} + T_\mathrm{e} \left( 1 - e^{-b_3 \left( \frac{\nu}{\nu_\ast} \right)^{-2}} \right),
\label{eq:T_F}
\end{equation}
}

where the effective spectral index is modeled as

\[
\alpha'(\nu) = \alpha + b_1 + b_2 \log\left( \frac{\nu}{\nu_\ast} \right).
\]

The parameters $b_1$ and $b_2$ account for overall and frequency-dependent deviations from a pure power law, respectively. The ionospheric term is modeled as radiative transfer through a layer with optical depth $b_3$ and electron temperature $T_\mathrm{e}$.

To constrain these parameters, we apply a Bayesian inference framework to the publicly released EDGES spectrum from \citet{bowman}. {This integrated spectrum was obtained by the EDGES collaboration after calibrating raw voltage spectra to antenna temperature using an absolute calibration scheme that includes laboratory and field measurements of the receiver, antenna reflection coefficient, and losses. The publicly available spectrum includes corrections for frequency-dependent antenna beam effects based on electromagnetic beam simulations combined with a diffuse sky model, as well as data quality cuts, RFI excision, and time averaging over hundreds of hours.}\\

We statistically validate each model using the discrepancy method of \citet{chaparro}, which compares {the discrepancy between observations, model-generated-synthetic data and the model expectation values to test for over- or under-fitting. After validation, this procedure} yields a posterior distribution over the foreground and ionospheric correction parameters, from which we generate a family of statistically validated spectra that can be stochastically sampled and applied to our sky-beam simulations.

\subsubsection{Bayesian extraction of foreground parameters}

Bayesian inference provides a systematic framework for constraining model parameters using observational data. In this work, we apply it to the publicly released EDGES spectrum \citep[]{bowman} to infer the posterior distribution of parameters in our foreground and ionospheric model. The likelihood function quantifies the probability of observing the data under a specific set of model assumptions, including those associated with the absence or presence of a cosmological 21\,cm absorption feature.

We adopt a Gaussian likelihood function defined over measurements of sky temperature $T_{\mathrm{sky}}$ as a function of frequency $\nu$, given by,

{\footnotesize
\begin{equation}
\ln p(T_{\mathrm{EDGES}}|\boldsymbol{\theta}) =
-\frac{1}{2}\sum_{n=1}^{N_\nu}
\left[
\frac{\left(y_n - T_{\mathrm{sky}}(\nu_n;\boldsymbol{\theta})\right)^2}{\sigma^2}
+ \ln\!\left(2\pi\sigma^2\right)
\right],
\label{eq:verosimilitud}
\end{equation}
}

{where $n$ labels the frequency channels of the EDGES spectrum, $y_n$ is the measured sky temperature at frequency $\nu_n$, and $N_\nu$ is the total number of frequency channels, $\sigma^2$ is the dispersion of the data with respect the parameter distribution, $\theta$ is a vector of the physically motivated parameters of each model} and $T_{\mathrm{EDGES}}$ is a vector of the observed data. Therefore, eq.~(\ref{eq:verosimilitud}) will return the probability that the initial parameter vector $\theta$ fits the sky model used from the data measured and reported by the EDGES project. We created uniform a priori probability functions for each parameter, reflecting the uncertainty about the $\theta$ parameters provided for the likelihood function,

\begin{equation}
    \ln P_{\mathrm{prior}}(\theta_i) = \left\{ 
        \begin{aligned}
            \mathrm{constant} \quad \mathrm{if} \quad & \theta^\text{min}_{i}\leq\theta_i\leq \theta^\text{max}_{i}\\
            0 \quad \mathrm{if} \quad & \theta_i\notin[\theta_i^\text{min},\theta_i^\text{max}]
        \end{aligned}
    \right.
    \label{eq:prior}
\end{equation}

where $\theta^\text{min}_{i}$ and $\theta^\text{max}_{i}$ determine the extreme values of the intervals of each parameter, and $\theta_i$ are the parameters initially set for the data fitting model. Our choice of uniform priors here is {intended to avoid explicitly favoring any specific foreground model structure within the adopted parameterization, and are tailored to finding closest-to-optimal Bayesian $p$-values for all models.}. The posterior function is obtained by applying Bayes' rule in eq. (\ref{eq:verosimilitud}) and (\ref{eq:prior}) of each parameter in a single probability function. The bounding values for the parameter priors used in each model can be found in Figs.~14,16,17.

{\footnotesize
\begin{equation}
    \ln P_{\mathrm{posteriori}} (T_{\mathrm{EDGES}}|\theta_i) = \ln P_{\mathit{prior}}(\theta_i) + \ln P(T_{\mathrm{EDGES}}|\theta_i) .
	\label{eq:posteriori}
\end{equation}
}

Eq.~(\ref{eq:posteriori}) represents the updated probability of the EDGES signal description model parameters, according to the intervals established in the a priori functions for each $\theta_i$.

\begin{figure*}
    \centering
    \includegraphics[width=0.8\textwidth]{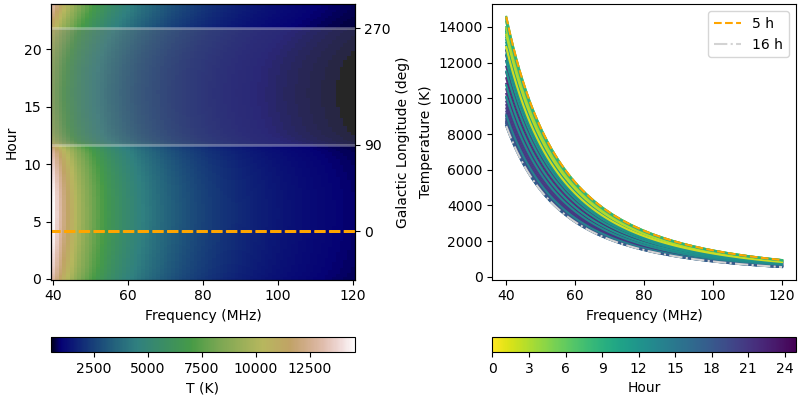}
    \caption{\textbf{Left:} Simulated dynamic spectra for an isotropic, achromatic antenna at the ALMA site on 2021-06-14. Color scale indicates antenna temperature as a function of frequency and time. The dashed orange line marks when Galactic longitude $0^\circ$ transits the zenith; the shaded region corresponds to when the zenith lies between 90$^\circ$ and 270$^\circ$ in Galactic longitude. \textbf{Right:} Spectral slices at representative times: 05:00 UTC (orange dashed) and 16:00 UTC (gray dashed), showing the range of sky brightness observed throughout the day.
}
    \label{fig:DS_ISO}  
\end{figure*}

\subsubsection{Statistical validation}

To assess the goodness of fit of the physical model to the EDGES data, we apply the discrepancy-based statistical validation framework introduced by \citet{chaparro}. This method, not employed in the original EDGES analysis, is based on a Bayesian test for model adequacy and can help determine a) whether a 21\,cm cosmological signal is supported by the data and b) best-fit foreground parameter distributions. For this, we generated a synthetic data array and used a measure of discrepancy $D$ between the data and the model-derived expected values for the same data,

\begin{equation}
D(\boldsymbol{\theta}_i) =
\sum_{j=1}^{M}
\left(
\sqrt{S^{\mathrm{obs}}_j}
-
\sqrt{S^{\mathrm{rep}}_j(\boldsymbol{\theta}_i)}
\right)^2 ,
\label{eq:discrepancias}
\end{equation}

{where $j$ indexes the frequency channels, $S^{\mathrm{obs}}_j$ is the observed
model-data residual variance in channel $j$,
$S^{\mathrm{rep}}_j(\boldsymbol{\theta}_i)$ is the corresponding residual variance
computed from a posterior predictive realization generated using parameter set
$\boldsymbol{\theta}_i$. The sum runs over the $M$ frequency channels included in
the validation statistic. This form corresponds to a Freeman-Tukey discrepancy statistic, comparing observed and posterior predictive variances on a per-channel basis.} For each extracted parameter $\theta_i$, it is then possible to compare the simulated discrepancy to the observed discrepancy. If the model adequately represents the data, the synthetic datasets generated from the model will match the observed data both in their values and statistical variability, resulting in similar simulated and observed discrepancies. Model adequacy is quantitatively assessed via the Bayesian $p$-value, defined as the fraction of simulations in which the synthetic discrepancy exceeds the observed discrepancy. 

In this Bayesian validation framework, the $p$-value quantifies how typical the observed discrepancy is compared to discrepancies from synthetic datasets generated from the fitted model. Specifically, a $p$-value close to 0.5 indicates that the model and data are statistically consistent, with discrepancies from observed data neither systematically larger nor smaller than expected. In contrast, extreme values below 0.05 or above 0.95 imply a poor model fit, reflecting systematic under- or over-estimation of observational uncertainties, respectively. For example, a Bayesian $p$-value near 1.0 indicates that the observed residuals are systematically smaller than the model’s predictive scatter, suggesting that the assumed noise or uncertainty has been overestimated or that the model fits data variations too closely, failing to capture the expected statistical variability. {We should note that allowing the observational uncertainty to vary freely does not permit arbitrarily large noise values, since models with predictive variances much larger than those observed are rejected by the posterior predictive validation as underfitting.}

{We emphasize that this statistical validation differs from a statistical evidence-based or information-criterion model comparison. In the posterior predictive framework used here, a model is considered adequate only if it can reproduce both the spectral structure and the intrinsic variance of the observed data. Models that produce posterior predictive realizations with variance systematically larger than observed are interpreted as underfitting, while models that produce variance systematically smaller than observed are interpreted as overfitting, even if their best-fit residuals appear visually acceptable. Extreme Bayesian p-values close to 0 or 1 therefore indicate model inadequacy in this framework, and not statistical preference.}

\begin{figure*}
    \centering
    \includegraphics[width=0.8\textwidth]{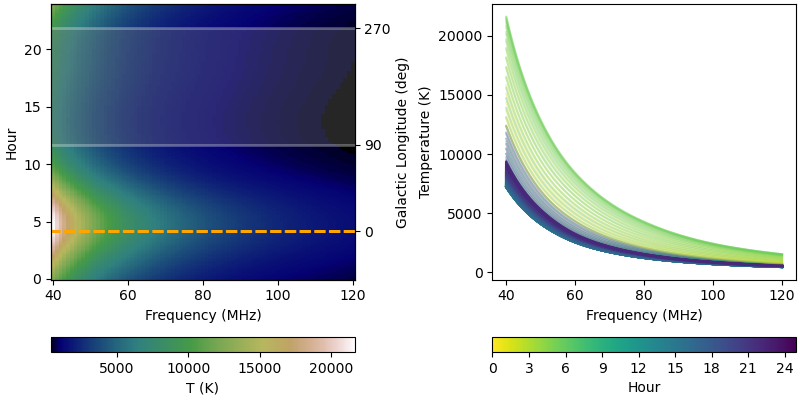}
    \caption{\textbf{Left:} Simulated dynamic spectra for the opt-blade antenna. Tick marks and shading follow the same convention as in Figure~\ref{fig:DS_ISO}. \textbf{Right:} Spectral slices at selected times, with the shaded region indicating periods when the zenith Galactic longitude lies between 90$^\circ$ and 270$^\circ$.
}
    \label{fig:DS_MIST}  
    \end{figure*}

\begin{figure}
    \centering
    \includegraphics[scale=0.55]{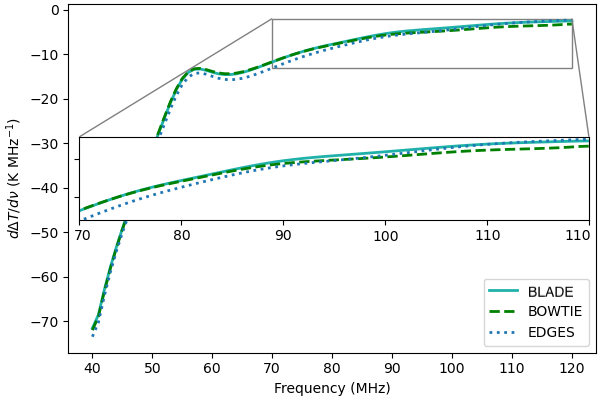}
    \caption{Derivative of the residuals between the mean spectra of the isotropic, achromatic antenna and those of the opt-blade, opt-bowtie, and EDGES antennas. The plot is restricted to the 70-120\,MHz range, corresponding to the frequency band of the reported absorption feature in \citet{bowman}.
}
    \label{fig:DerRes_MIST}  
\end{figure}

\section{Results}
\label{Res}

\subsection{Chromaticity of antenna temperature with opt-blade and opt-bowtie vs. EDGES antenna models}

We simulated beam-convolved dynamic spectra across the 40-120\,MHz frequency range, encompassing the band relevant for global 21\,cm signal detection, with a spectral resolution of 1\,MHz. Observations were simulated over a 24-hour period on an arbitrary reference date (2021-06-14), divided into 6-minute intervals. In the dynamic spectra shown throughout this section, the antenna temperature is plotted as a function of frequency and time over a 24-hour sidereal day, with shading used to indicate intervals when the zenith points away from the brightest regions of the Galactic plane (i.e., Galactic longitudes between 90 and 270 degrees). These masked time windows represent periods of lowest expected foreground contamination. All spectra are simulated for the ALMA site (latitude -23.03 degrees) on the reference date 2021-06-14.

As a baseline, we first simulated the sky as observed by an ideal isotropic and achromatic antenna, located at the ALMA site. The resulting dynamic spectrum is shown on the left-hand side of Fig.~\ref{fig:DS_ISO}, and representative spectral slices extracted at specific times are shown on the right-hand side.

Sky temperature varies significantly throughout the day, primarily due to the changing position of the Galactic plane relative to the antenna beam. The highest temperatures, exceeding 13{,}000\,K, occur when the Galactic center (at 0$^\circ$ Galactic longitude) transits the zenith, as indicated by the orange line in Fig.~\ref{fig:DS_ISO}. In contrast, when the zenith lies between 90$^\circ$ and 270$^\circ$ Galactic longitude, the antenna views the radio-quietest region of the sky. These intervals are highlighted by grey shaded regions in Figs.~\ref{fig:DS_ISO} and \ref{fig:DS_MIST}. We further compare spectra at representative times: 05:00 UTC (dashed orange line), when the Galactic center dominates the beam, and 16:00 UTC (dashed white line), when the beam points toward a relatively structureless region.

The same frequency and time resolution were applied in simulations of the EDGES, opt-blade, and opt-bowtie antennas. While their dynamic spectra appear qualitatively similar, we perform a more detailed comparison using spectral slices. We focus initially on the opt-blade antenna (Fig.~\ref{fig:DS_MIST}) and identify time windows consistent with our foreground mask (i.e., zenith Galactic longitudes between 90$^\circ$ and 270$^\circ$). We observe that the opt-bowtie and EDGES simulations exhibit comparable behavior. For each antenna, we compute the mean integrated spectrum over the selected observation window, simulating the effect of long integration during a full observing run.

To evaluate antenna chromaticity, we analyze residuals between the mean spectra and the isotropic reference. As shown in Fig.~\ref{fig:DerRes_MIST}, the opt-bowtie antenna most closely reproduces the isotropic baseline, indicating minimal chromatic distortion. Additionally, the opt-blade antenna exhibits the lowest standard deviation in the 40--80\,MHz band, the same frequency range as the absorption feature reported by \citet{bowman}.

These results demonstrate that the PSO-optimized designs proposed by \citet{restrepo}, particularly the opt-bowtie configuration, exhibit favorable spectral smoothness and chromaticity performance. The opt-bowtie antenna is the best-performing design across our {designs}, especially if the absorption signal lies near 78\,MHz, as suggested by the EDGES result. These findings show the promise of these optimized antennas for future global 21\,cm experiments.

\begin{table}
\centering
\begin{tabular}[t]{lccc}
\hline
&40-120 MHz&40-80 MHz&80-120 MHz\\
\hline
Opt-blade&16.11&18.17&1.50\\
Opt-bowtie&16.00&18.31&1.26\\
EDGES&16.41&18.23&1.82\\
\hline
\end{tabular}
\caption{Standard deviation (in K MHz$^{-1}$) for the derivative of the residuals of the opt-blade, opt-bowtie, and EDGES antennas compared to the isotropic antenna.}
\label{tab:STD_MIST}  
\end{table}

\begin{figure*}[t]
            \centering
            \includegraphics[scale=0.8]{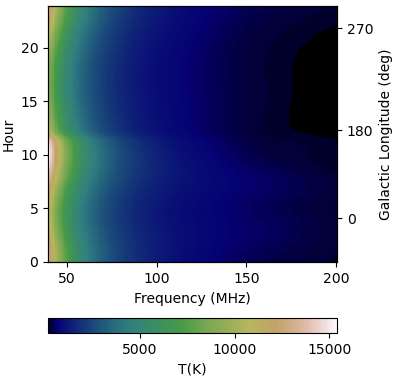}
            \caption{{Simulated dynamic spectrum for a representative site-testing monopole antenna at the ALMA location. The color scale shows antenna temperature as a function of frequency and time. The remaining monopole configurations exhibit qualitatively identical spectra and are therefore not shown.}}
            \label{fig:Dynamic Spectra MINI MIST}  
\end{figure*}

\subsection{Chromaticity of monopole antenna variants for site testing}

To assess the performance of our proposed monopole antenna variants for site testing, we compute the dynamic spectra for the ALMA site for the initial (monopole-1), winged (monopole-2) and wingless (monopole-3) antennas (Figure \ref{fig:Dynamic Spectra MINI MIST}). We also use a simulation of an isotropic, achromatic antenna as a comparison model.

\begin{figure}[htbp!]
            \centering
            \includegraphics[scale=0.5]{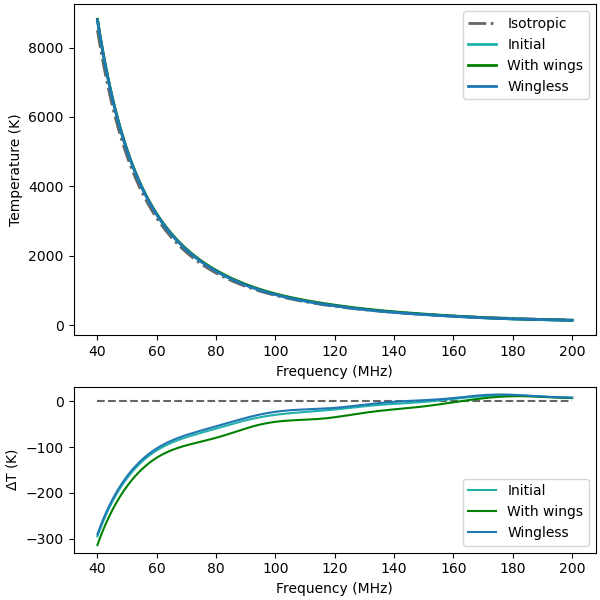}
            \caption{Top: Spectra at 06:00 local time (Chile) for the three monopole antennas and the isotropic reference antenna, showing differences in spectral response. Bottom: Residuals with respect to the isotropic antenna.
}
            \label{fig:Spectra17h}  
\end{figure}

Fig.~\ref{fig:Spectra17h} shows the spectra for the time when the Galactic centre is not above the horizon, corresponding roughly to 17h (pointing to a mostly structureless sky). To better appreciate the subtle differences in antenna performance, Fig.~\ref{fig:Spectra17h}  also shows the residuals of the initial (monopole-1), winged (monopole-2), and wingless (monopole-3) antennas compared to the isotropic antenna. We see the similarities in the operation of the initial (monopole-1) and wingless (monopole-3) antenna, while the winged antenna (monopole-2) differs more from the radiation pattern of the isotropic antenna. With the help of our radio sky models, these monopole site testing antennas could be used alongside the main 21 cm cosmology experiment antenna for foreground model validation and extraction.

\begin{figure}[htbp!]
            \centering
            \includegraphics[scale=0.5]{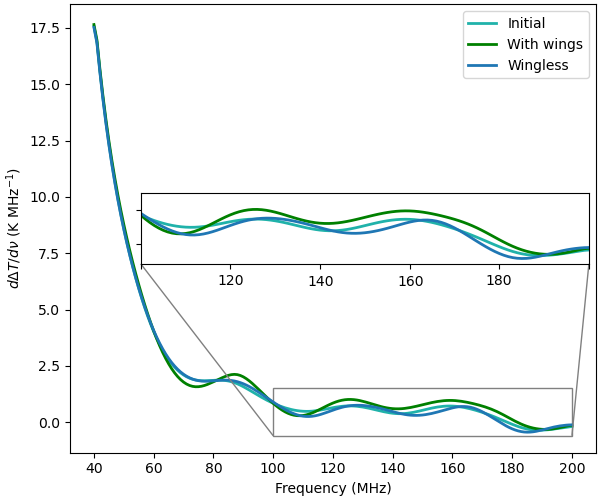}
            \caption{Derivative of the residuals between the monopole antennas and the isotropic reference antenna, showing differences in spectral smoothness across frequency.
}
            \label{fig:DerRes}  
\end{figure}

Figure~\ref{fig:DerRes} demonstrates that taking the derivative of the residuals provides enhanced sensitivity to antenna chromaticity. While all three monopole designs exhibit similar behavior below 60\,MHz, distinct differences emerge at higher frequencies. The initial monopole design (monopole-1) and the wingless configuration (monopole-3) display comparable spectral responses, whereas the winged variant (monopole-2) shows significantly larger deviations in the highlighted frequency range. These results indicate that the optimized monopole-1 and monopole-3 designs offer the most favorable chromatic performance and are thus the most suitable candidates for low-cost 21\,cm cosmology site testing and foreground model validation.

\subsection{Spectral Smoothness and Residual Analysis}

To quantify the chromaticity of each antenna, we calculate the residual spectrum defined as the difference between each antenna’s mean integrated spectrum and the ideal isotropic reference antenna spectrum. We then compute two metrics: the standard deviation of these residuals (in K), which quantifies the overall amplitude of chromatic structure, and the standard deviation of the frequency derivative of these residuals (in K MHz$^{-1}$, capturing more subtle frequency-dependent variations in the spectral shape. Throughout this analysis, these two distinct metrics will be explicitly distinguished by their units to avoid confusion.

These metrics were evaluated over the 40–120\,MHz band and separately within the 40–80\,MHz sub-band, which includes the frequency range of the claimed EDGES absorption feature. In the full band, the opt-bowtie antenna exhibited the smallest residual deviation ($\sigma \approx 15\,\mathrm{mK}$) and the lowest derivative standard deviation, indicating high spectral smoothness. The opt-blade antenna showed slightly higher deviation ($\sigma \approx 20\,\mathrm{mK}$), while the EDGES antenna exhibited the largest residual fluctuations across the band ($\sigma \approx 35\,\mathrm{mK}$). In the lower sub-band (40–80\,MHz), both the opt-blade and monopole-3 designs performed especially well, with derivative RMS values under $5\,\mathrm{mK\,MHz^{-1}}$. These quantitative results reinforce the conclusion that the PSO-optimized antennas, particularly the opt-bowtie and monopole-3, are best suited for low-chromaticity applications in global 21\,cm signal experiments.

\subsection{Bayesian model validation and posterior-based foreground correction}

To evaluate the consistency of different physical interpretations of the EDGES spectrum, we make a Bayesian model comparison using four configurations of a physically motivated foreground and ionospheric radiative transfer model. These models are based on the parameterization described in Sect.~\ref{stoch} and eq.~\ref{eq:T_F}, which captures synchrotron spectral index variation and ionospheric emission and absorption. Each model is tested against the publicly released EDGES data using a consistent Bayesian framework.

Our analysis is based on a physically motivated model \citep{hills2018} that incorporates both the spectral structure of the Galactic sky and the radiative transfer through the ionosphere. Instead of relying on polynomial foreground subtraction \citep{bowman}, we use a generative model with parameters grounded in astrophysical considerations. We then apply Bayesian inference via MCMC to capture uncertainties in the sky, ionosphere, and instrumental response, and use the discrepancy test mentioned above \citep{chaparro} to assess consistency between the model and the data. 

{The Bayesian parameter inference was performed using the affine-invariant ensemble sampler implemented in the \texttt{emcee} package \citep{foreman}. For each model configuration, we initialized an ensemble of 1k walkers and evolved each walker for 25k steps. The initial positions of the walkers were drawn from the prior distributions. An initial burn-in phase was discarded based on autocorrelation time estimates, and the remaining chains were thinned by a factor of 100 before being flattened for analysis. }

{To assess whether the Markov chains have converged to the target posterior distribution, we apply the $\hat{R}$ diagnostic criterion \citep{gelman,vehtari} using the \texttt{ArviZ} library \citep{arviz}. This tool evaluates convergence by comparing the between-chain variance to the within-chain variance. When convergence is achieved, both variances should be statistically indistinguishable \citep[]{cita8}.}

{The $\hat{R}$ statistic, also known as the potential scale reduction factor, is defined as}

\begin{equation}
    \hat{R} = \frac{\hat{V}}{\hat{W}},
    \label{eq:rhat}
\end{equation}

{where $\hat{W}$ is the within-chain variance and $\hat{V}$ is an estimate of the posterior variance based on pooled rank traces. Values of $\hat{R}$ approaching 1 indicate convergence, while values significantly above 1 suggest that additional sampling is required. Following standard practice, we consider chains to be well converged if $\hat{R} < 1.01$ for all model parameters.} Under this framework, we find that the EDGES absorption profile is not statistically supported when using physically plausible assumptions.

We compare the following four model variants:

\begin{enumerate}
    \item \textbf{Absorption + Variable Error}: The full model including a 21\,cm absorption profile, with variable observational uncertainty treated as a free parameter. This approach reflects the fact that EDGES did not report a definitive error model and allows for underestimated residual structure.
    
    \item \textbf{Absorption + Fixed Error}: The same physical model including the absorption feature, but assuming a fixed error value based on the best estimate of the previous model.
    
    \item \textbf{No Absorption + Variable Error}: A model with no cosmological 21\,cm signal component, allowing variable error to assess whether the observed spectral structure can be accounted for entirely by foregrounds and ionospheric effects.
    
    \item \textbf{No Absorption + Fixed Error}: A purely foreground and ionosphere model without a 21\,cm feature, using a fixed noise level based on the best estimate of the previous model.
\end{enumerate}

\begin{figure*}
    \centering
    \includegraphics[width=0.8\textwidth]{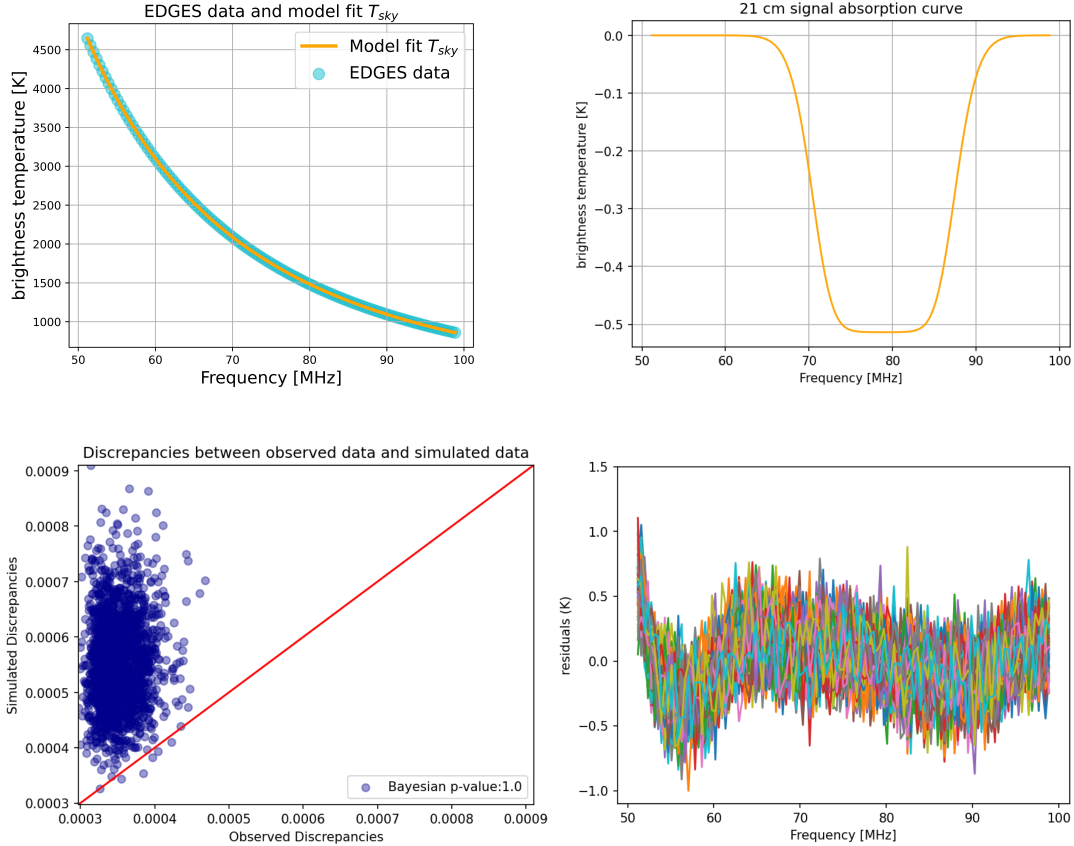}
    \caption{\textbf{Top left:} Sky model $T_{\mathrm{sky}}$ including foregrounds, ionospheric effects,  21\,cm absorption, and variable error (Model 1). \textbf{Top right:} Modeled 21\,cm absorption profile. \textbf{Bottom left:} Discrepancy comparison between synthetic spectra and the EDGES data; Bayesian $p$-value = 0.98. \textbf{Bottom right:} Residuals between the observed EDGES data and the model realization. {Each curve shows posterior predictive realizations, including both parameter uncertainty and the stochastic component of the noise model.}
}
    \label{fig:img_modelo-con_absorcion_sig_variable}  
\end{figure*}

These configurations allow us to isolate the effects of (i) including or excluding a cosmological absorption signal, and (ii) assuming fixed versus free observational uncertainty. For each model, we compute the posterior distribution of parameters using Markov Chain Monte Carlo (MCMC) sampling, assess statistical consistency with the data using the discrepancy method of \citet{chaparro}, and compare the resulting Bayesian $p$-values.

This model comparison framework allows us to statistically assess whether the EDGES spectrum requires a 21\,cm absorption component and whether such an interpretation is sensitive to the assumed error model. Additionally, it provides validated foreground and ionospheric parameter posteriors from the measured data, which can then be used to generate stochastically corrected sky models for antenna–beam convolution. This allows for a consistent propagation of astrophysical uncertainty into global 21\,cm signal simulations.

\subsubsection{Model 1: Absorption with Variable Error}

In this configuration, we include a 21\,cm absorption feature in the physically motivated model of foreground and ionospheric emission, as described in eq.~\ref{eq:T_F}. To account for the fact that the EDGES data release does not specify a full error model, we introduce a free noise scaling parameter that allows the observational uncertainty to vary. This enables the model to accommodate uncharacterized residual structure that may arise from instrumental or calibration effects.

\begin{figure*}
    \centering
    \includegraphics[width=0.8\textwidth]{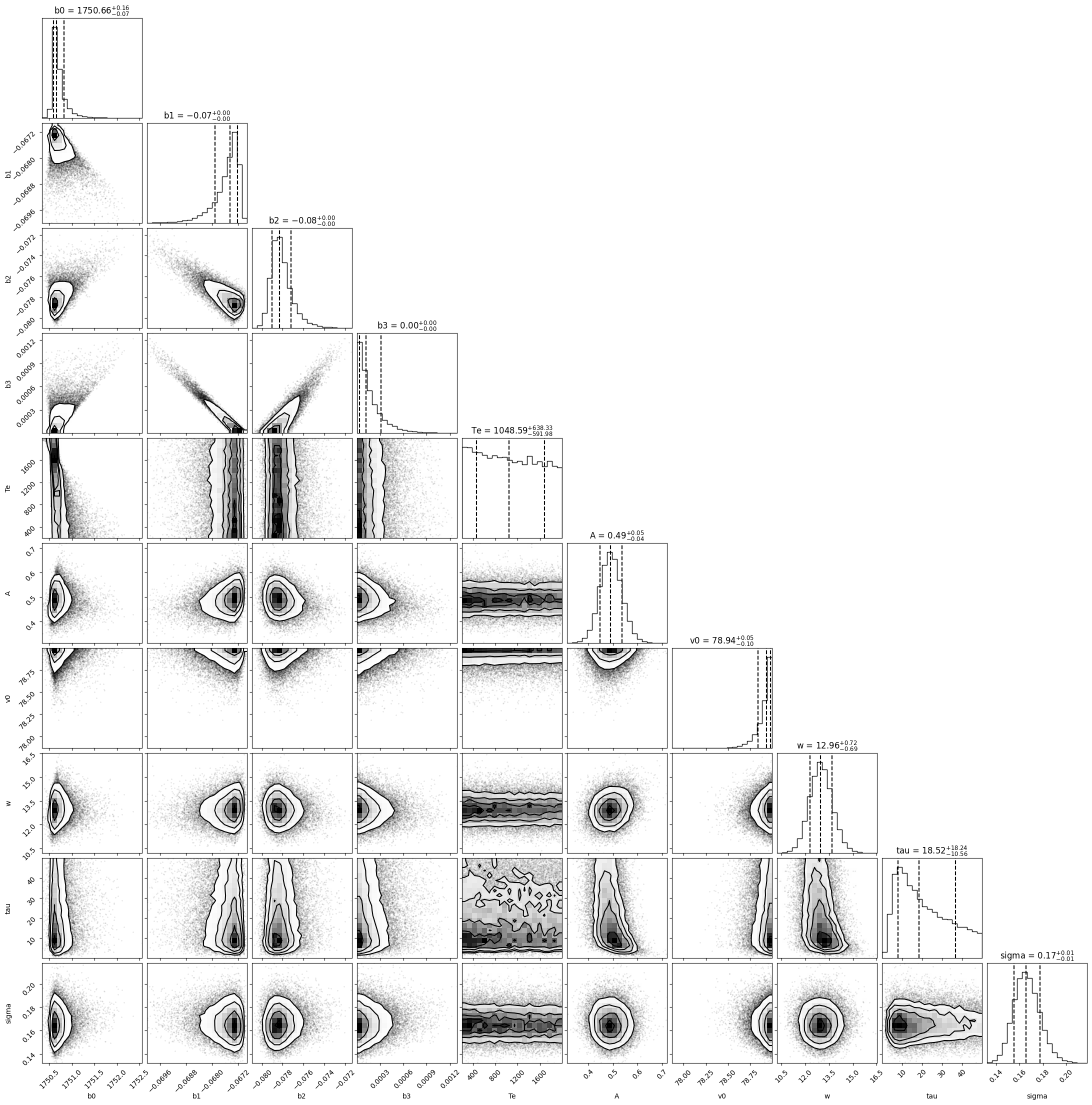}
    \caption{Posterior distributions of the parameters in the physical foreground model $T_{\mathrm{sky}} = T_F + T_{21}$ using Model 1 (including absorption using variable error). }
    \label{fig:cornetplot_modelo_con_absorcion_sigma_variable}  
\end{figure*}

The simulated spectrum combines the foreground and ionospheric components with a cosmological 21\,cm absorption profile centered near 78\,MHz, as reported by the EDGES collaboration \citep{bowman}. Using a Monte Carlo sampling procedure, we generate posterior distributions for the model parameters via Bayesian inference, employing a Gaussian likelihood and astrophysically informed priors. The corner plot of the resulting posteriors is shown in Fig.~\ref{fig:cornetplot_modelo_con_absorcion_sigma_variable}.

Figure~\ref{fig:img_modelo-con_absorcion_sig_variable} shows the best-fit model spectrum compared to the EDGES data. While the model reproduces the general shape of the reported absorption feature, {although the model clearly seems to prefer higher values for the central frequency than that reported by EDGES, and yields a $\sigma\sim0.2$~K.}. The discrepancy analysis yields a Bayesian $p$-value of $\sim1$, indicating that the model is statistically inconsistent with the observed spectrum \citep{chaparro}. We therefore reject this configuration: although flexible enough to fit the data, it fails to produce simulated realizations that are statistically supported by the observations.

\begin{figure*}
    \centering
    \includegraphics[width=0.4\textwidth]{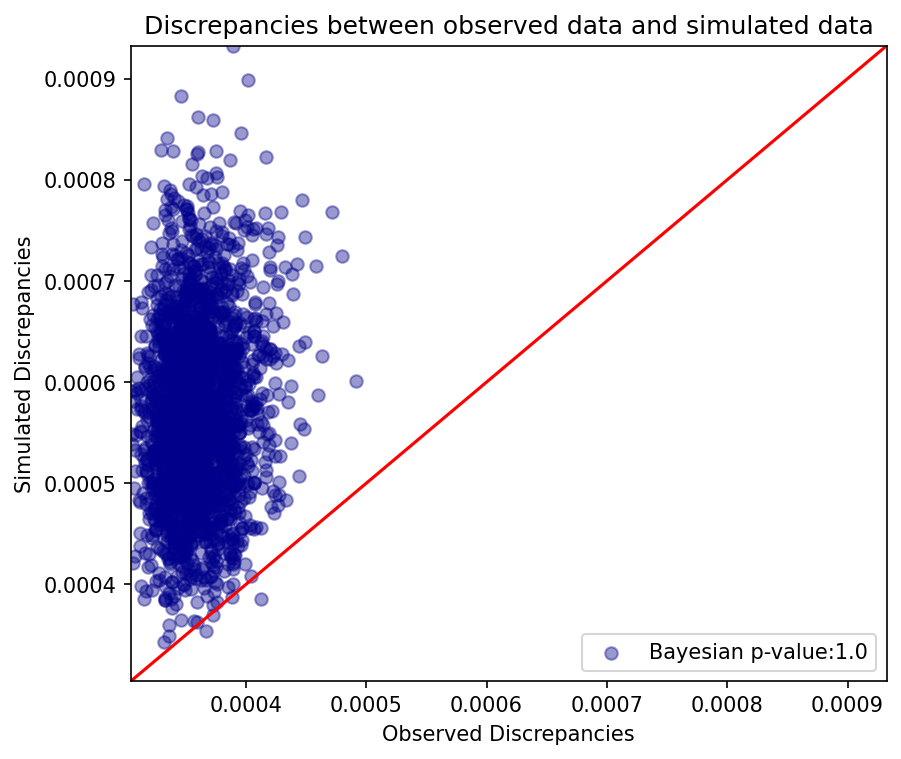}
        \includegraphics[width=0.4\textwidth]{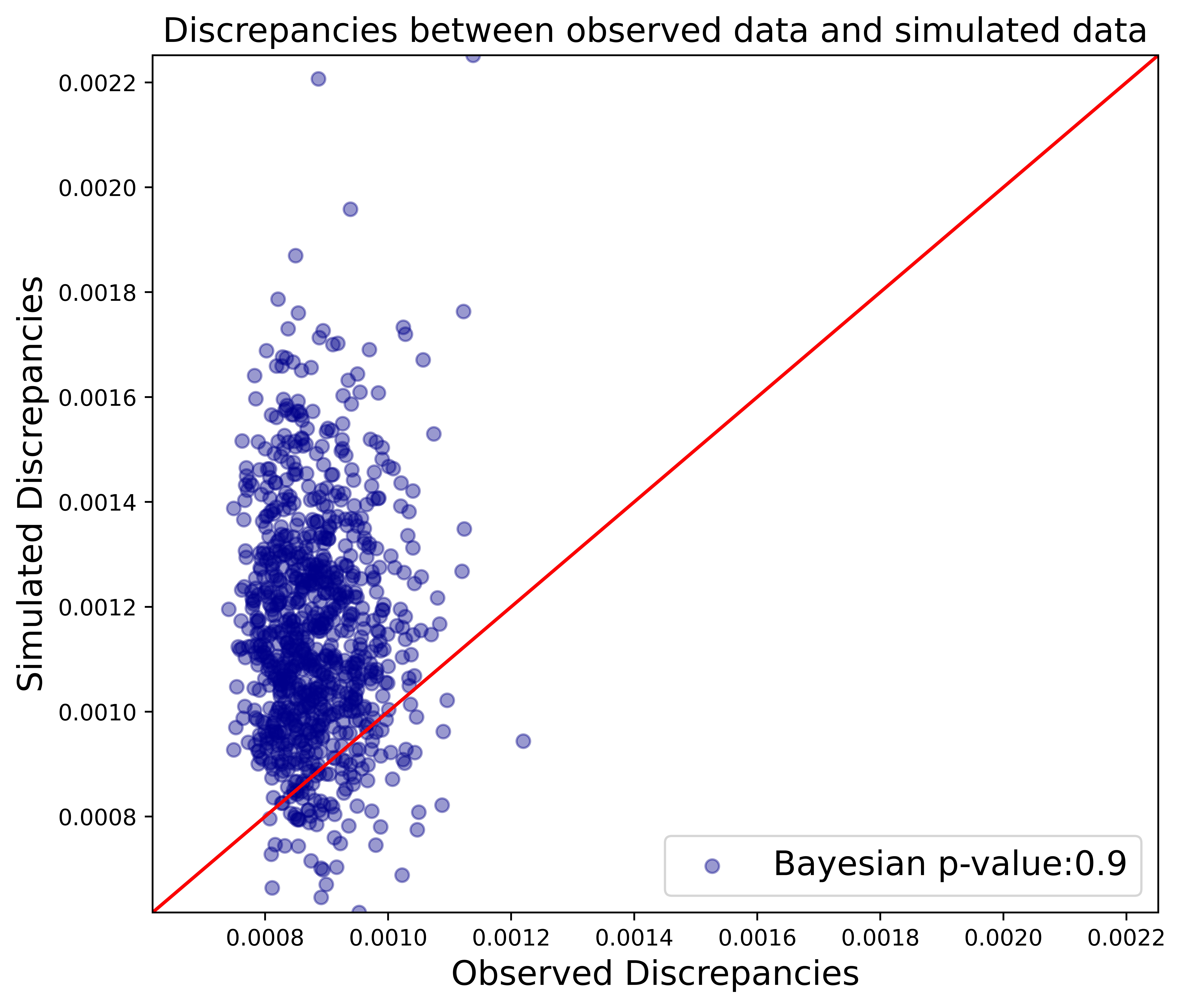}

    \caption{{\textbf{Left:} Discrepancy comparison between synthetic spectra for a model with absorption and a fixed sigma (Model 2) and the EDGES observations, yielding a Bayesian $p$-value of 1.0, which means that this model is rejected for overfitting. \textbf{Right:} Discrepancy comparison between simulated spectra and the EDGES observational data for the no-absorption, variable-error model (Model 3). The Bayesian $p$-value of 0.9 indicates a barely sufficient statistical agreement.}}
    \label{fig:img_modelo_con_absorcion_sig_fijo}  
\end{figure*}

\subsubsection{Model 2: Absorption with Fixed Error}

This model is identical to Model 1 in its inclusion of a physically motivated 21\,cm absorption profile and foreground–ionosphere emission, but assumes a fixed observational uncertainty {in order to inspect whether the model validity improves}. We fix the noise level to $\sigma = 0.17$\,K, corresponding to the median value obtained for the noise parameter in Model 1’s posterior distribution.

Using the same Bayesian framework, we compute the posterior distribution of the model parameters and assess the fit via the discrepancy-based statistical validation. While this model also reproduces the absorption feature reported by EDGES, the statistical analysis yields a Bayesian $p$-value of 1.0, indicating that the the posterior predictive distributionare also highly inconsistent with the observed data (Fig.~\ref{fig:img_modelo_con_absorcion_sig_fijo}, left). As in Model 1, we reject this configuration based on its inability to generate statistically consistent realizations of the EDGES spectrum.

Taken together, the results of Models 1 and 2 demonstrate that no statistically valid fit to the EDGES data can be achieved under the assumption of a physically motivated 21\,cm absorption signal. {Regardless of whether} the error model is allowed to vary freely, the posterior draws fail to reproduce the observed spectrum in a statistically meaningful way.

This shows that no combination of signal and uncertainty parameters within this framework can account for the observed spectrum. While earlier analyses such as that of \citet{hills2018} have remarked on the degeneracy and instability of the EDGES signal under parametric foreground fitting, to our knowledge, {posterior predictive validation has not been systematically used as the primary criterion to assess absorption-inclusive versus absorption-free fits to the publicly released EDGES spectrum within this class of physically motivated models}. Our results demonstrate that the reported spectrum does not support a cosmological 21\,cm absorption interpretation, neither in shape, nor in amplitude, nor in statistical validity.

\begin{figure*}
    \centering
    \includegraphics[width=0.8\textwidth]{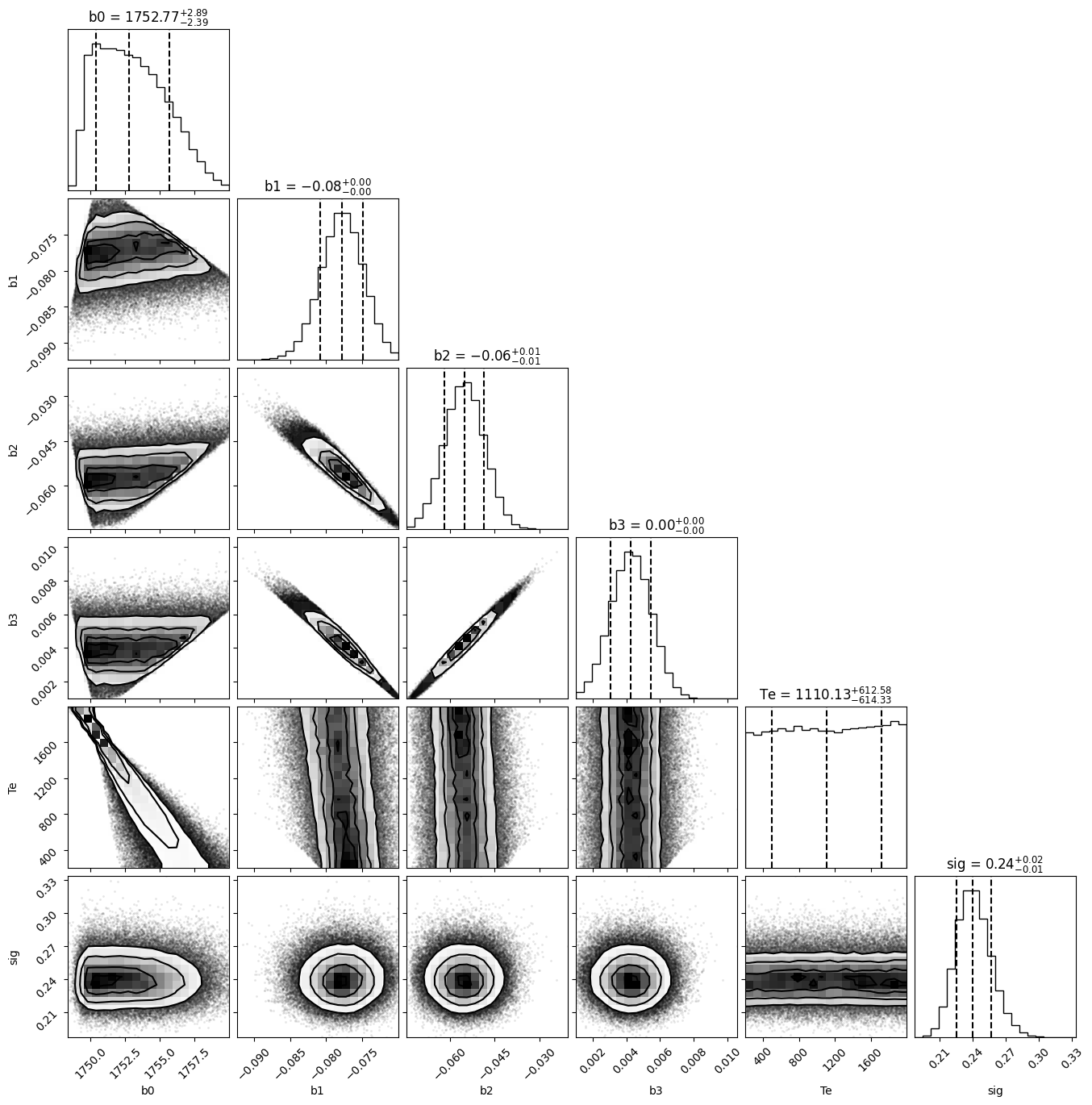}
    \caption{Corner plot of posterior distributions for the no-absorption, variable-error model (Model 3). The parameters remain within physically motivated bounds, and the systematic error stabilizes around $\sigma = 0.24$\,K.
}
    \label{fig:cornet_sin_abs_sig_var}  
\end{figure*}

\subsubsection{Model 3: No Absorption with Variable Error}

In this configuration, we remove the 21\,cm absorption profile from the physically motivated model, focusing instead on fitting only the foreground and ionospheric components. The observational uncertainty is treated as a free parameter, allowing us to assess how well the EDGES spectrum can be explained without invoking a cosmological signal.

Figure~\ref{fig:img_modelo_con_absorcion_sig_fijo} (right) shows that this model yields a good fit to the EDGES data. The discrepancy analysis returns a Bayesian $p$-value near the upper end of the acceptable range in \citet{chaparro}, indicating marginal but sufficient statistical agreement between the model and the data. The posterior distributions, shown in Fig.~\ref{fig:cornet_sin_abs_sig_var}, remain well constrained within the bounds imposed by the priors.

{The free stochastic error parameter $\sigma$ stabilizes at near 0.24~K, which is consistent with the \citet{hills2018} best fit for $\log\sigma(\text{K})\approx-1.6$.  Although the EDGES collaboration reports residual levels from calibration tests at the 10–300 mK scale depending on load temperature, these reflect systematic effects rather than stochastic noise.  This suggests that the model can reproduce the observed spectrum with physically plausible parameter values, rejecting a 0.5~K cosmological absorption signal as implausible.}

\subsubsection{Model 4: No Absorption with Fixed Error}

Motivated by the outcome of Model 3, we test a fourth configuration in which the 21\,cm absorption feature is again excluded, but the observational uncertainty is fixed at $\sigma = 0.24$\,K  {(median $\sigma$ from Model 3) in order to check whether this leads to a more stable model}. We can then to examine whether removing the noise degeneracy affects the robustness of the parameter inference.

\begin{figure*}[t]
    \centering
    \includegraphics[width=0.8\textwidth]{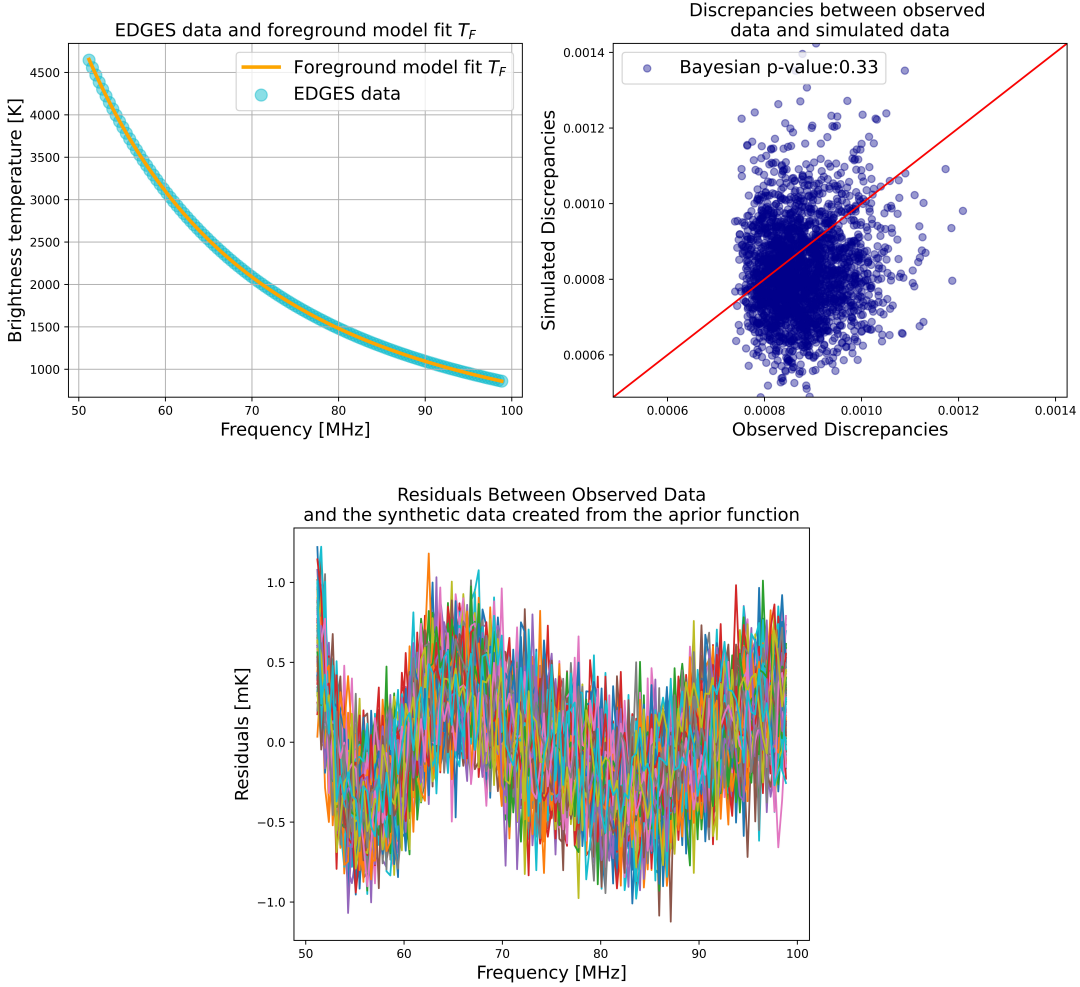}
    \caption{\textbf{Top left:} Best-fit of the physically motivated model without 21\,cm absorption, incorporating Galactic foregrounds and ionospheric effects, with a fixed 0.24 K rms error (Model 4). \textbf{Top right:} Discrepancy comparison between simulated spectra and EDGES data, showing improved agreement without an absorption component. \textbf{Bottom:} Residuals between the modeled and observed EDGES spectra.}
    \label{fig:model_sin_abs_sig_fix}  
\end{figure*}

\begin{figure*}
    \centering
    \includegraphics[width=0.8\textwidth]{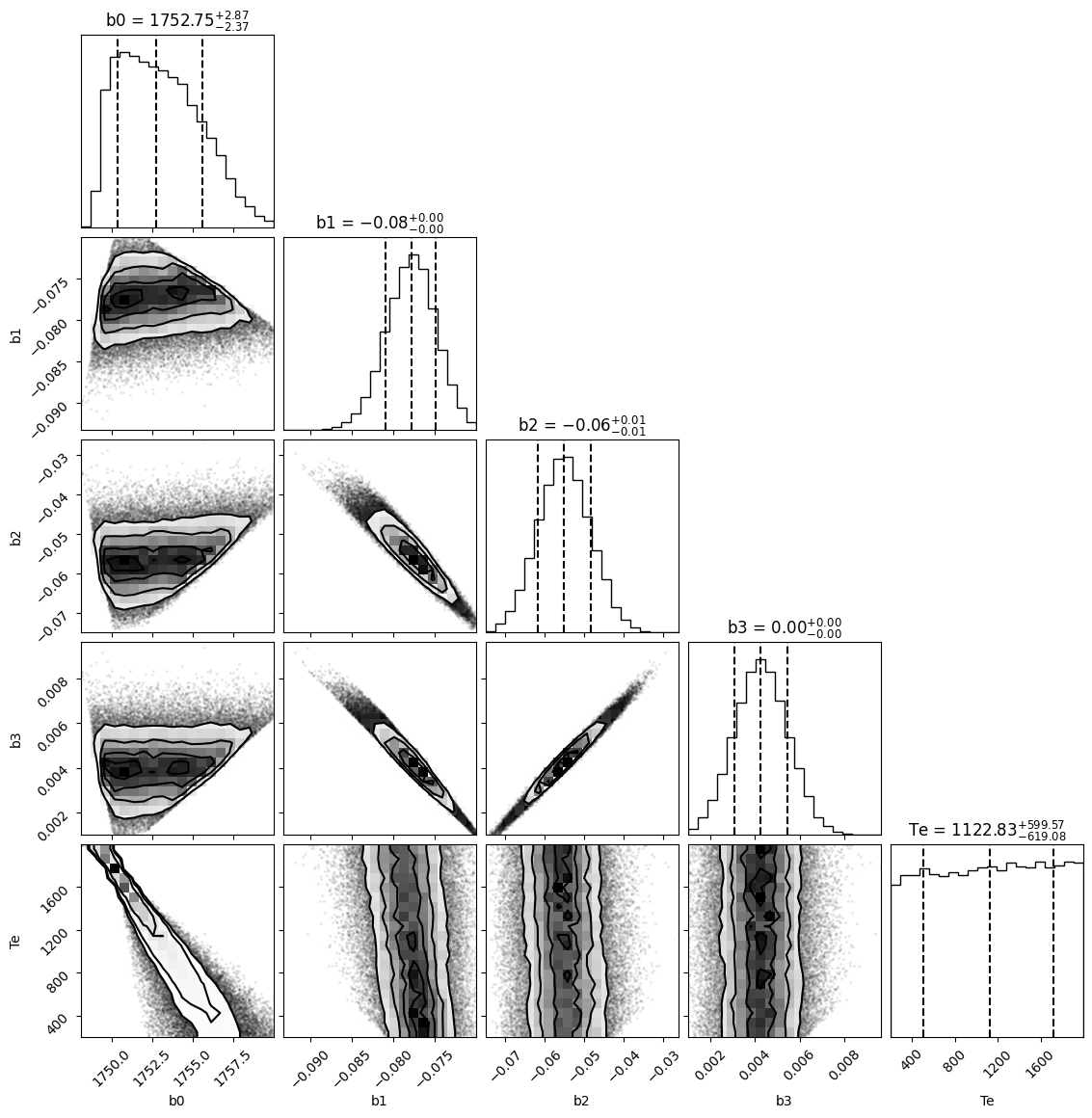}
    \caption{Corner plot of posterior distributions for the no-absorption model with fixed $\sigma = 0.24$\,K (Model 4). Compared to the variable-error case, this configuration yields tighter parameter constraints and more stable correlations among $b_0$, $b_1$, $b_2$, $b_3$, and $T_e$, reflecting a more rigid but statistically consistent fit. 
}
    \label{fig:cornetplot_modelo_sin_absorcion_sigma_fijo}  
\end{figure*}

As shown in Fig.~\ref{fig:model_sin_abs_sig_fix}, this model provides an excellent fit to the EDGES spectrum. The residuals between the model and data are unbiased, with a typical deviation below 1\,mK. The Bayesian $p$-value of 0.3 confirms that the model is statistically consistent with the observations and not overfit. The corner plot in Fig.~\ref{fig:cornetplot_modelo_sin_absorcion_sigma_fijo} demonstrates that all model parameters ($b_0$, $b_1$, $b_2$, $b_3$, and $T_e$) are well sampled and remain within physically plausible ranges defined by the priors.

We conclude that Model 4 provides a physically and statistically robust explanation of the EDGES data without requiring a cosmological 21\,cm absorption feature. Therefore, it can be useful in extracting physically meaningful parameters related to foreground emission and ionospheric radiative transfer. Even in the absence of a cosmological 21\,cm absorption signal, {the EDGES data can lead to a model that} can be used to characterize the spectral structure of the radio sky and the impact of ionospheric effects.

\subsection{Impedance matching analysis of residuals}

\begin{figure}
    \centering
    \includegraphics[width=\linewidth]{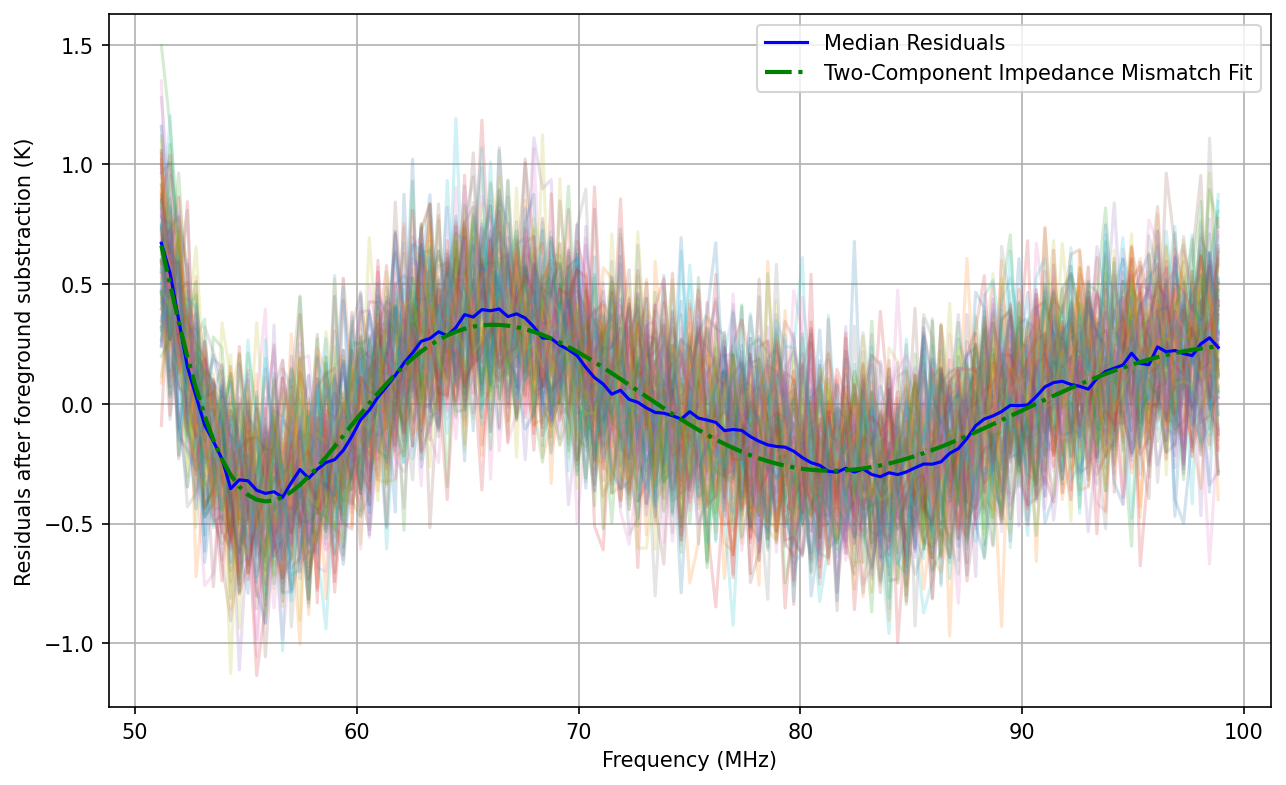}
    \caption{{Residual analysis for the EDGES low–band data after subtraction of our
five–parameter foreground model.  Thin coloured lines show 500 random noise
realisations of the residuals ({Model 4}) spanning
50–100\,MHz; their envelope illustrates the 240\,mK scatter expected from
thermal noise.  The thick blue curve is the median residual, revealing a
tapered, log-periodic structure.  The green dash–dotted line is the best-fit
two-component impedance-mismatch model
(equation~\ref{eq:two_comp}), which reproduces the chromatic
ripple and reduces the post-foreground removal rms\ to 39\,mK. }}
    \label{fig:res_anal}
\end{figure}

{Figure \ref{fig:model_sin_abs_sig_fix} shows residual draws from our statistically preferred model (Model 4) with respect to the original EDGES data. These residual draws $R(\nu)$ show a spectral structure that could be hiding a purported absorption signal of 0.5~K. However, upon inspection the residual shows a tapered log-periodic structure that could be well explained by a chromatic effect induced by impedance mismatch \citep{sims} rather than by a single absorption dip. An impedance mismatch between the dipole feed and the low–noise amplifier is modelled as a single-path reflection with complex gain,}
\begin{equation}
G(\nu)=1+\Gamma\,e^{2\pi i\nu\tau},
\label{eq:gain}
\end{equation}
{where $\tau$ is the round–trip delay and $\Gamma$ is the reflection coefficient.
Expanding $G$ to first order in $|\Gamma|\ll1$ and keeping the real
part yields the additive term}
\begin{equation}
\Delta T(\nu)=A\Bigl(\tfrac{\nu}{\nu_\ast}\Bigr)^{-\beta}\!
\cos\!\bigl[2\pi\tau\nu+\phi\bigr],
\label{eq:ripple_linear}
\end{equation}
{with $A\simeq\Gamma\, T_{\rm sky}(\nu_\ast)$.
For the narrow fractional bandwidth $50-100$~MHz we set
$\nu=\nu_\ast e^{x}$ ($x=\ln\nu/\nu_\ast$,
$|x|\lesssim0.4$) and expand $e^{x}=1+x+\tfrac12x^{2}+\cdots$.  Neglecting
$\mathcal{O}(x^{2})$ terms and absorbing the constant phase,
equation~(\ref{eq:ripple_linear}) becomes}
\begin{equation}
R_i(x)=A_i\,{e}^{-\beta_i x}\cos\!\bigl(k_i x+\phi_i\bigr),
\quad
k_i\equiv2\pi\tau_i\nu_\ast,
\label{eq:ripple_log}
\end{equation}
{i.e. an exponentially tapered periodic function of $\ln\nu$. Although we attempted a residual fit with a single impedance component, a better fit was obtained with two components (Figure \ref{fig:res_anal}), so the full template is}
\begin{equation}
R_{\rm tot}(x)=\sum_{i=1}^{2}
A_i\,{\rm e}^{-\beta_i x}\cos\!\bigl(k_i x+\phi_i\bigr).
\label{eq:two_comp}
\end{equation}

{Table~\ref{tab:fit} summarises the best–fit parameters obtained from a
non–linear least–squares optimisation of equation~(\ref{eq:two_comp}) to the
median residual. The two–component model reduces the rms residual from
$240$~mK (after foreground subtraction) to $39$~mK over 50–100~MHz. . }

\begin{table}[ht]
\centering
\caption{{Best–fit impedance-mismatch derived chromatic parameters for the EDGES residual}}
\label{tab:fit}
\begin{tabular}{lcccccc}
\hline
Component & $A_i$ [K] & $\beta_i$ & $k_i$ & $\phi_i$ &
$\tau_i^{\rm (a)}$ [ns] & $|\Gamma_{0,i}|^{\rm (b)}$ \\
\hline
1 & $+0.292$ & $-0.583$ & $15.016$ & $1.869$ &
$31.9$ & $1.7\times10^{-4}$ \\
2 & $-0.004$ & $-14.443$ & $20.436$ & $4.901$ &
$43.4$ & $2.3\times10^{-6}$ \\
\hline
\multicolumn{7}{l}{%
\footnotesize
${}^{\rm (a)}~\tau_i=k_i/(2\pi\nu_\ast)$ with $\nu_\ast=75$~MHz. \qquad
${}^{\rm (b)}~|\Gamma_{0,i}|=|A_i|/T_{\rm sky}(\nu_\ast)$, 
$T_{\rm sky}(75{\rm~MHz})=1749$~K.}
\end{tabular}
\end{table}

{The first component yields $\tau_1\simeq32$~ns, whereas the second component yields a $\tau_1\simeq43$~ns. This may be a weak second reflection or a beam–edge diffraction mode. The coefficients $|\Gamma_{0,i}|$ in Table~\ref{tab:fit} are {\em effective}
values at the detector input.  Because the signal is attenuated on the
forward and reverse traversals of the lossy front end, the relationship
between the antenna–terminal mismatch $|\Gamma_{\rm ant}|$ and the residual
amplitude is}
\begin{equation}
|\Gamma_{0,i}| \;=\; |S_{\rm fwd}|^{\,2}\,|\Gamma_{\rm ant}|
\;\xrightarrow{|S_{\rm fwd}|\simeq0.07}\;
\bigl(5\times10^{-3}\bigr)\,|\Gamma_{\rm ant}|.
\label{eq:gamma_eff}
\end{equation}
{A measured feed mismatch of $|\Gamma_{\rm ant}|\approx0.15$
($\mathrm{RL}\simeq-16.5$~dB) could appear in the residual
at the $\approx5\times10^{-3}$ level (–46~dB).  The additional suppression
introduced by the five–term foreground polynomial lowers the observable
amplitude by roughly an order of magnitude, bringing the surviving ripple
down to the $10^{-4}$ range found in the fit. This is consistent with
laboratory $S_{11}$ measurements of EDGES-type blade antenna scaled models \citep{restrepo}.}

\subsection{Stochastically corrected sky-beam model}

Following the Bayesian model comparison in the previous sections, we use the validated posterior from Model 4 (no absorption, fixed error) to generate a stochastically corrected sky model. This approach incorporates the effects of synchrotron spectral curvature and ionospheric radiative transfer into our beam-convolved simulations using the physically motivated parameterization of \citet{hills2018}. The model includes foreground spectral index variation and an ionospheric transfer function, constrained by the EDGES observational data \citep{bowman}.

From the posterior samples, we construct a family of statistically consistent foreground and ionospheric spectra. These realizations reflect the range of physically plausible sky temperatures compatible with the data, and we use them to propagate uncertainty through the antenna–sky convolution.

\begin{figure}[t]
    \centering
    \includegraphics[scale=0.55]{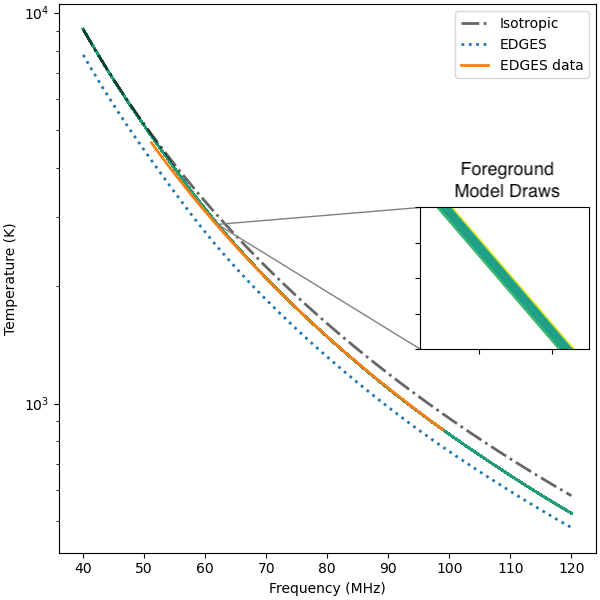}
    \caption{Stochastically corrected model realizations of the mean sky spectra, incorporating radio foreground and ionospheric transfer parameters inferred from the EDGES data. Shown alongside are the uncorrected simulations for the EDGES antenna and an ideal isotropic antenna.
}
    \label{fig:bayesian}  
\end{figure}

Figure~\ref{fig:bayesian} shows a subset of these realizations for the EDGES antenna, plotted alongside the uncorrected EDGES data and an ideal isotropic reference. The corrected spectra reproduce the overall structure of the observed sky temperature and remain within the range allowed by the data. The amplitude of the stochastic corrections is typically $\lesssim 0.5$\,K, significantly smaller than the variations introduced by beam chromaticity. This confirms that, under validated foreground and ionospheric conditions, instrumental chromaticity remains the dominant source of spectral contamination in global 21\,cm signal experiments.

\subsection{Latitude dependence of dynamic spectra and implications for site selection}

\begin{figure}[t]
    \centering
    \includegraphics[scale=0.5]{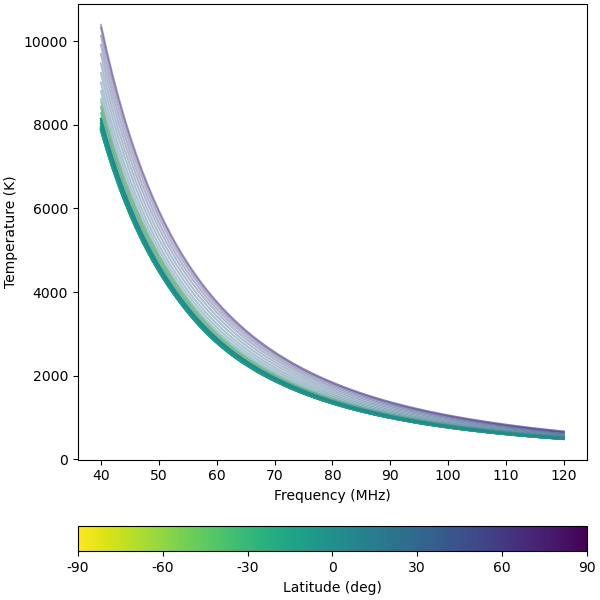}
    \caption{Mean sky spectra for different observing latitudes, averaged over times when the zenith Galactic longitude lies between 90$^\circ$ and 270$^\circ$. The lowest-temperature (highest-opacity) spectra correspond to mid-latitudes between $-40^\circ$ and $+5^\circ$, indicating optimal foreground suppression.
}
    \label{fig:DSGLats}  
\end{figure}

To evaluate how observing latitude affects sky temperature and foreground structure in global 21\,cm experiments, we simulated dynamic spectra for the opt-blade antenna at latitudes spaced by 5°, ranging from the South Pole to the Arctic. At each site, we calculated the mean antenna temperature over a sidereal day, including only the time intervals when the zenith pointed away from the brightest parts of the Galactic plane (i.e., Galactic longitudes between 90° and 270°). Figure~\ref{fig:DSGLats} shows the resulting mean spectra.

The simulations indicate that the lowest sky temperatures occur between latitudes of approximately $-40^\circ$ and $+5^\circ$. In this range, Earth's rotation enables the antenna to observe quieter sky regions for substantial portions of the day. The Galactic center periodically drops below the horizon, which allows foreground suppression through time masking. The ALMA site at $-23^\circ$ lies within this favorable zone, but several other locations, such as proposed future CANTAR sites in Colombia also fall within this latitude range and offer similarly advantageous sky access for low foreground intensity observations.

By contrast, high-latitude sites such as the Antarctic plateau (below $-60^\circ$) experience persistent visibility of the Galactic center. The minimal sky variation over a sidereal cycle leads to higher average sky brightness and limits the effectiveness of time-domain foreground filtering. Similarly, northern sites above $+60^\circ$ are affected by near-constant visibility of bright radio sources such as the Perseus arm and Cassiopeia A \citep{arms, morales}, which remain close to the zenith throughout the day.

While these results indicate that mid-latitude sites are optimal for science-grade observations, the Antarctic offers an exceptionally clean and stable RFI environment. This makes it an ideal location for calibration, hardware validation, and end-to-end pipeline testing for the CANTAR initiative under controlled conditions. Through the Programa Radioastronómico Antártico Colombiano (PRAC), we have completed three Antarctic deployments to date. These campaigns have measured the local RFI environment, characterized soil dielectric properties relevant to ground-plane coupling, and performed in situ antenna response measurements (Rodríguez et al, in prep).

The ongoing PRAC deployments will support aspects of the CANTAR experimental workflow, including beam pattern validation, soil–antenna interaction modeling, and development of foreground subtraction techniques. This work lays the foundation for robust, systematics-aware measurements at future science-grade sites.

\begin{figure*}[]
    \centering
    \includegraphics[width=0.5\textwidth]{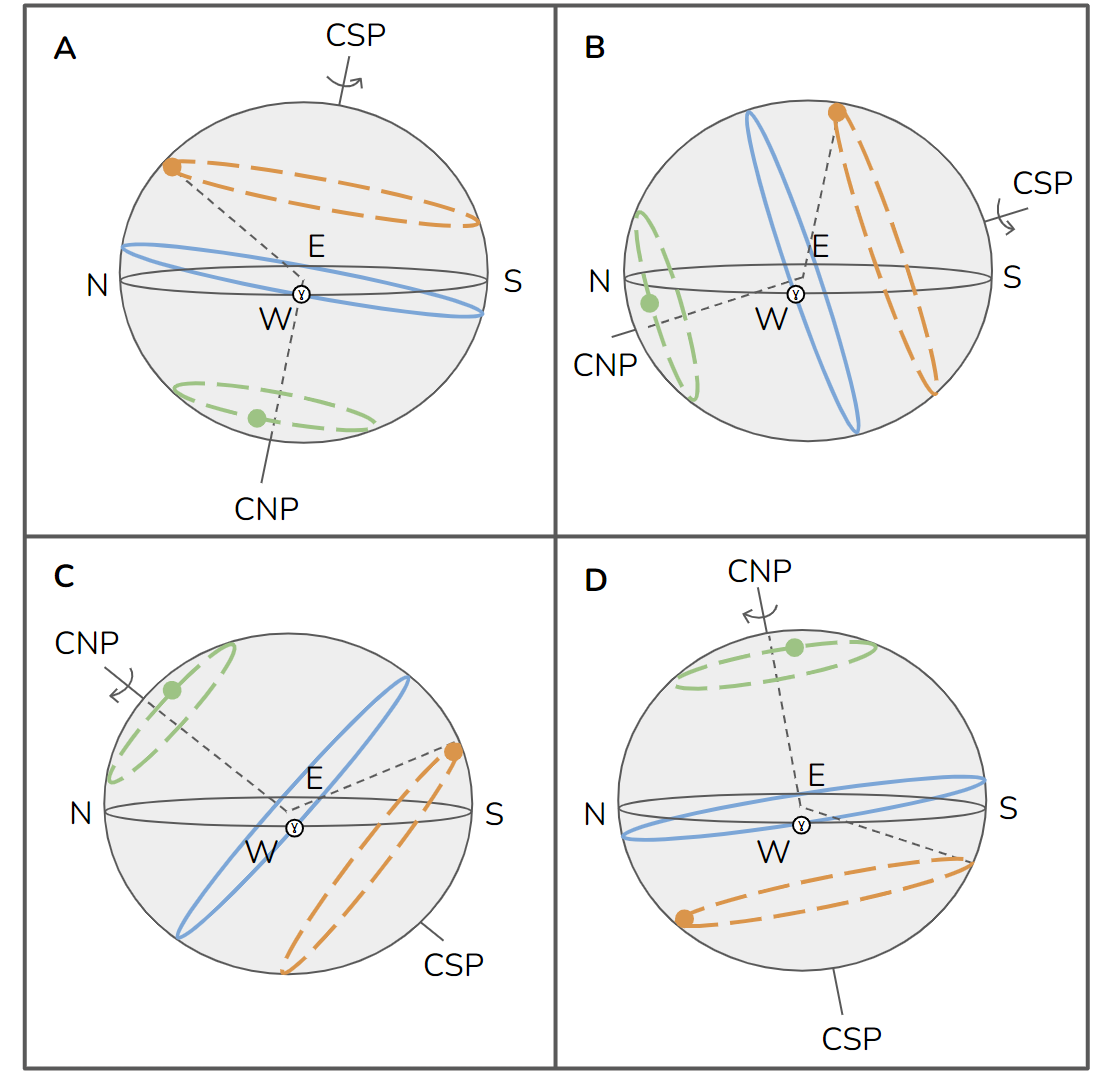}
    \caption{Sky visibility at different latitudes, shown for a fixed sidereal time with the vernal point to the west. The orange dot marks the Galactic center (Sagittarius A*), and the green dot marks Cassiopeia A. Dashed lines trace their diurnal motion; the blue curve shows the celestial equator. {The geometry is shown in the local horizon frame, which is the natural frame for describing beam-weighted sky temperatures and the time-variation of the foreground.} \textbf{A.} Southern polar latitudes ($-90^\circ$ to $-60^\circ$): the Galactic center remains above the horizon. \textbf{B.} Mid-southern latitudes ($\sim$$-20^\circ$): both sources rise and set, allowing time windows with minimal contamination. \textbf{C.} Northern mid-latitudes ($\sim$$+40^\circ$): Cassiopeia A is frequently visible. \textbf{D.} Northern polar latitudes: Cassiopeia A remains above the horizon throughout the day.}
    \label{fig:skydrawing}  
\end{figure*}

\section{Conclusions}
\label{Conc}


We have developed a simulation and analysis framework for global 21\,cm experiments as part of the Colombian Antarctic Telescopes for 21\,cm Absorption during Reionization (CANTAR) project, combining beam–sky convolution, antenna optimization, Bayesian foreground modeling, and site evaluation. Particle swarm–optimized antenna designs \citep{restrepo} show significantly reduced chromaticity compared to the EDGES blade dipole {across the full band considered. The performance differences between the optimized designs are incremental rather than qualitative,} with the opt-blade performing {slightly better} at 40–80\,MHz and the opt-bowtie {yielding the lowest overall chromaticity across the} 40–120\,MHz range. Among monopole designs, the wingless and unmodified variants show the lowest chromaticity, making them suitable for low-cost site testing and foreground validation.

Applying a novel Bayesian model validation approach to the EDGES data,  {We apply a posterior predictive consistency test that evaluates whether the publicly released EDGES spectrum is statistically consistent with an absorption-inclusive model. We find that only models excluding a 21\,cm signal yield statistically consistent fits, and that the absorption feature can not be statistically validated. This approach does not assess the general ability of this framework to recover an injected 21 cm signal under idealized conditions, although we are currently working on such an extension to our current framework.}

{Furthermore, the residual structure that could be hiding a $\sim0.5$\,K absorption feature can also be explained by two weak reflection paths in the front-end.  Modelling the chromatic response as the sum of two exponentially tapered log-periodic terms reduces the post–foreground removal rms\ from 240\,mK to 39\,mK across 50–100\,MHz, leaving no statistically significant evidence for the originally reported 21-cm absorption spectral features. The effective reflection coefficients ($|\Gamma_{0,1}|\approx1.7\times10^{-4}$ and $|\Gamma_{0,2}|\lesssim10^{-5}$) are fully consistent with previous $S_{11}$ measurements once round-trip attenuation and foreground-polynomial suppression are accounted for. } 

{We therefore conclude that under our framework, the EDGES low-band spectrum, when corrected for these modest impedance mismatches, shows no compelling evidence for a global 21\,cm absorption at $z\sim17$; future experiments must control similar chromatic systematics below the $10^{-4}$ level to reach sub-100\,mK sensitivity.}

{Although absorption-inclusive models are not consistent with the data, within our framework we are able to construct stochastically corrected spectra from an absorption-removed, statistically validated foreground model that we are able to use to generate sky temperature models for antenna design, observation planning, site testing, and low-chromaticity foreground characterization using our proposed antenna prototypes}. 

Latitude-dependent sky simulations reveal that mid-latitude sites ($-40^\circ$ to $+5^\circ$) offer optimal foreground suppression. While not suitable for final science operations, our Antarctic deployments under the Programa Radioastronómico Antártico Colombiano (PRAC) program will provide a platform for low-RFI calibration, beam validation, and pipeline testing. Therefore, the CANTAR experiment can be steered toward a two phase approach: an Antarctic phase for calibration and validation, and a mid-latitude science phase optimized for statistically robust foreground suppression and signal extraction. Our work demonstrates that our two-phase strategy for the CANTAR global 21\,cm experiment is viable and compares well to other previous and current initiatives in the field of 21-cm cosmology observations.

 \section*{acknowledgments} 

GC and PC-R received support from Proyecto CODI 2024-73250, Vicerrectoría de Investigación, Universidad de Antioquia. JU and JG received support from the Joven Investigador 2021, 2024 program (respectively), Universidad de Antioquia. GC and JR-F acknowledge support from Programa Antártico Colombiano, Comisión Colombiana del Océano, Fuerza Aeroespacial Colombiana, Armada de Colombia, Instituto Antártico Chileno, Ejército de Chile, Armada de Chile, Fuerza Aérea de Chile for their support during our Antarctic Expeditions. Finally, we thank Ricardo Bustos for his valuable insight and comments.

\vspace{5mm}









\bibliography{sample631}{}
\bibliographystyle{aasjournal}



\end{document}